\documentclass[12pt]{article}
\usepackage{amsmath,amsfonts, amssymb, braket, bbold}
\usepackage{tikz}
\usetikzlibrary{trees,er,snakes,shapes,mindmap}
\usepackage{titlesec}
\titleformat{\section}
 {\normalfont\fontfamily{bch}\fontsize{13pt}{16pt}\bfseries\color{black}}
{\thesection}{1em}{}
\titleformat{\subsection}
 {\normalfont\fontfamily{bch}\fontsize{12pt}{16pt}\bfseries\color{black}}
{\thesubsection}{1em}{}
\def \beq  {\begin{equation}}
\def \eeq  {\end{equation}}
\def \beqar {\begin{eqnarray}}
\def \eeqar {\end{eqnarray}}
\allowdisplaybreaks
\def\sqr#1#2{{\vcenter{\vbox{\hrule height.#2pt
\hbox{\vrule width.#2pt height#1pt \kern#1pt
\vrule width.#2pt}\hrule height.#2pt}}}}

\def\L {{\cal L}}

\def\vx {{\vec x}}

\def\vf {{\varphi}}

\def\Tr {{\rm Tr}}
\def \tr {{\rm tr}}

\def\bD {\bar{D}}

\def\vx {{\vec x}}

\def\del {\partial}

\def\bz {{\bar{z}}}

\def\A {{\cal A}}

\def\D {{\cal D}}

\def\H {{\cal H}}
\def\I {{\cal I}}
\def \L {{\cal L}}
\def\M{{\cal M}}

\def\vf {{\varphi}}

\def\half{\textstyle{1\over 2}}

\mathchardef\mhyphen="2D
\begin{document}
\fontfamily{bch}\fontsize{12pt}{16pt}\selectfont
\def \CMP {{Commun. Math. Phys.}}
\def \PRL {{Phys. Rev. Lett.}}
\def \PL {{Phys. Lett.}}
\def \NPBProc {{Nucl. Phys. B (Proc. Suppl.)}}
\def \NP {{Nucl. Phys.}}
\def \RMP {{Rev. Mod. Phys.}}
\def \JGP {{J. Geom. Phys.}}
\def \CQG {{Class. Quant. Grav.}}
\def \MPL {{Mod. Phys. Lett.}}
\def \IJMP {{ Int. J. Mod. Phys.}}
\def \JHEP {{JHEP}}
\def \PR {{Phys. Rev.}}
\def \JMP {{J. Math. Phys.}}
\def \GRG{{Gen. Rel. Grav.}}
\begin{titlepage}
\null\vspace{-62pt} \pagestyle{empty}
\begin{center}
\vspace{1.3truein} {\large\bfseries
Higher dimensional quantum Hall effect and the analog of}\\
~\\
{\large\bfseries $W_\infty$-algebra}
\vskip .1in
{\large\bfseries ~}\\
~\\
~\\
{\sc Dimitra Karabali$^{a,c}$,  V.P. Nair$^{b, c}$}\\
\vskip .2in
{\sl $^a$Physics and Astronomy Department,
Lehman College, CUNY\\
Bronx, NY 10468}\\
\vskip.1in
{\sl $^b$Physics Department,
City College of New York, CUNY\\
New York, NY 10031}\\
 \vskip .1in
\begin{tabular}{r l}
{\sl E-mail}:&\!\!\!{\fontfamily{cmtt}\fontsize{11pt}{15pt}\selectfont 
dimitra.karabali@lehman.cuny.edu}\\
&\!\!\!{\fontfamily{cmtt}\fontsize{11pt}{15pt}\selectfont vnair@ccny.cuny.edu}\\
\end{tabular}
\vskip .5in

\centerline{\large\bf Abstract}
\end{center}
We show that Abelian and non-Abelian gauge transformations
are the analog of $W_\infty$ transformations for higher dimensional
quantum Hall effect.
The commutator anomaly and the extended algebra of such transformations on the edge modes of a droplet are obtained by purely topological arguments,
basically utilizing the two-cocycle in the descent procedure for
anomalies and using the fact that there is anomaly cancellation between the bulk and boundary actions. 
The method relies on the fact that bulk actions are easily constructed in general using the Dolbeault index theorem.
The resulting algebras are shown to agree with explicit edge mode calculations for cases 
where they are available. We also comment on the similarities and differences in the nature of these transformations
 between two and higher dimensions.

\end{titlepage}
\fontfamily{bch}\fontsize{12pt}{16pt}\selectfont
\pagestyle{plain} \setcounter{page}{2}
\section{Introduction}
The quantum Hall effect has been the subject of abiding research interest
for a few decades, both experimentally and theoretically \cite{QHE}.
Although much of the research has been on the two-dimensional
case, many features of this effect and the mathematical 
structures involved admit interesting
extensions to higher dimensions \cite{HZ}-\cite{everyone}.
Interestingly, QHE in higher dimensions
is also experimentally realizable
using the idea of synthetic dimensions \cite{exp}.
Edge excitations of a droplet of fermions is one of the
many interesting features of QHE.
It is perhaps easiest to characterize these for 
quantum Hall states with filling fraction equal to one.
Since inter-Landau level transitions are energetically disfavored for high values of the
magnetic field, one may consider the low energy dynamics as confined to a 
 particular Landau level, say, the lowest one. 
In two dimensions, the degeneracy of the level is given by the area of the sample in units of an elemental area defined by the magnetic length, 
with each elemental area corresponding to a single
quantum state. Since the exclusion principle forbids multiple occupancy, each fermion takes up an elemental unit of area. They also cluster towards the minimum of any additional confining potential, thus forming a droplet of area proportional to the number of fermions. This droplet behaves as a fluid in the limit of large number of fermions, it is also incompressible
by virtue of the exclusion principle.
Since the occupied area has to be preserved, the low energy excitations of this droplet will correspond to area-preserving diffeomorphisms of the droplet,
manifesting as chiral waves on the edge of the droplet.
Quantum theoretically, each fermion occupies a state in the single-particle Hilbert space $\H$. Excitations correspond to unitary transformations
on $\H$, modulo the invariances of the density matrix of the droplet.

The canonical or symplectic structure for the single-particle dynamics
 is proportional to the geometric
area, by virtue of projection to a single Landau level.
Therefore area-preserving diffeomorphisms correspond to canonical
transformations; in particular they can be viewed as $U(1)$-valued gauge transformations
of the symplectic potential or one-form.
In the quantum theory, they are therefore naturally elevated to
unitary transformations of the single-particle Hilbert space.

The edge excitations can be described by a chiral boson field,
in terms of which one can define a current operator which is
part of the generator of the $U(1)$ transformations mentioned above.
The key defining feature for the edge modes
 is that there is a central extension for the algebra
of currents given by a two-cocycle which is related to the anomaly structure
of the theory. The purpose of this paper is to show how this arises from the bulk action via
the descent method for anomalies, a method which can then be extended to
the higher dimensional cases.

It is useful to recall some of the key features 
familiar from two dimensions before proceeding to the case of arbitrary even dimensions. The transformation of interest is the $U(1)$ gauge transformation
of the electromagnetic vector potential, which is also accompanied by
the transformation
$\psi \rightarrow e^{i \theta} \psi$, $\psi^\dagger \rightarrow
e^{-i \theta} \psi^\dagger$ for the fermion field operators.
The canonical generator of this transformation on $\psi$, $\psi^\dagger$ is
the density operator $ \rho (\theta) = \int \theta \, \psi^\dagger \psi$,
with the commutation rule
$[\rho(\theta), \rho(\theta'] = 0$.
The fermion field operator restricted to the lowest Landau level
(LLL) has the expansion
\beq
\psi = \sum_k c_k \, \Psi_k (x), \hskip .3in
\psi^\dagger = \sum_k c^\dagger_k \, \Psi^*_k(x)
\label{intro1}
\eeq
where $\{\Psi_k\}$ are a basis of single-particle LLL wave functions and $c_k$,
$c^\dagger_k$ are the annihilation and creation operators.
With this restriction $\rho$ does not commute with itself for different test functions, but obeys the rule
\beq
[\rho (\theta), \rho(\theta')] = \rho( \{\!\{ \theta, \theta'\}\!\})
\label{intro2}
\eeq
where $\{\!\{ \theta, \theta'\}\!\}$ denotes the commutator of the
star-product of $\theta$, $\theta'$.
This is the $W_\infty$-algebra in two dimensions
\cite{IKS}-\cite{walg}.\footnote{The nature of the star-commutator is not relevant for now, so we do not display it here.
It is basically the star-commutator for the contravariant symbol,
as explained in \cite{vpn-berezin}.}
Notice that this is really the $U(1)$ gauge transformation
restricted to a particular Landau level.
One can also apply this transformation to the edge modes, it then
gives an extension to the current algebra on the edge.
This feature, we emphasize, will hold in general.
In higher dimensions, the
analog of $W_\infty$ will again be the $U(1)$ (and possible non-Abelian)
gauge transformations restricted to a particular Landau level.
Alternate interpretations such as area-preserving diffeomorphisms
will be dimension-dependent.

Returning to the general case, we start by showing, in the next section,
 how the commutator anomaly for the generators of the gauge transformation is related to a two-cocycle \cite{anomalies1}-\cite{anomalies3} which is in turn, related to the descent equations for anomalies \cite{anomalies3}.
We will then use the fact that any lack of gauge invariance
is canceled between the explicit variation of the bulk action and the
anomaly of the boundary theory
to relate the extension of the current algebra, or the commutator anomaly of the edge modes, to the bulk action.
The {\it raison d'\^etre} for this approach is that, 
while one can work out the extension of the current 
algebra by standard canonical quantization if an edge action is
available, 
an explicit expression 
for the edge action has been derived only for certain geometries
\cite{{KN2},{KN3},{Kar}}.
On the other hand, generally for complex manifolds of any even dimension, one can construct an effective action for the
bulk dynamics of the system, including fluctuations of the
gauge fields and the metric, using arguments based on the Dolbeault
index theorem \cite{{KN-dolb},{AKN}}. This is briefly reviewed in section 3.
So by relating the extension of the current algebra to the bulk action,
we can derive a general expression for any even dimension.
This is carried out in section 4.

The anomaly-descent part of our argument applies in general. It requires a bulk topological effective action whose gauge variation reproduces the boundary anomaly, but does not by itself require a complex structure. In the present paper, such bulk actions are constructed using the Dolbeault index, and this specialization assumes that the spatial manifold is complex. Quantum Hall systems on manifolds without a complex structure, such as the four-sphere with a non-Abelian instanton background, should instead be formulated using the appropriate Dirac-type index density. Once the corresponding bulk Chern-Simons action is identified, the same descent procedure determines the two-cocycle and hence the extension of the edge current algebra. A detailed implementation for such geometries is beyond the scope of the present work.

A few other points regarding the situation in arbitrary dimensions are worthy of comment. Primarily, the symmetry of interest is the $U(1)$ gauge transformation of the symplectic potential. This coincides with area-preserving diffeomorphisms in two dimensions because the symplectic two-form is proportional to the geometric area.
As mentioned before, the
expression of this symmetry
in the bulk is the $W_\infty$-transformation, with
the restriction to the edges giving the current algebra
\cite{IKS}-\cite{walg}.
In higher dimensions, the corresponding symmetries are again
$U(1)$ transformations which preserve the symplectic form.
The phase volume is preserved,\footnote{Up to a multiplicative constant, the phase volume is the same as the geometrically defined volume for 
a complex manifold with constant Abelian curvature, 
including manifolds such as $\mathbb{CP}^k$.} but the situation is more restrictive, the
two-form itself has to be preserved. So, in higher dimensions, the
analogs of the $W_\infty$ transformations do not correspond to an arbitrary volume-preserving transformation, but a restricted set. Secondly, in higher dimensions we also have the possibility of non-Abelian
background fields, and, correspondingly, non-Abelian gauge symmetries.
The edge dynamics will show an anomaly for these as well, leading to
a commutator anomaly for the corresponding currents.
So we have these additional current algebraic structures
in higher dimensions, going beyond the familiar
$W_\infty$ of the two-dimensional
situation. 
The full theory maintains the symmetry with cancellation of the
variation of the bulk action and the anomaly due to the boundary modes.
So we again use our arguments to work out the corresponding current
algebra in section 4. 
As mentioned earlier, actions for the edge modes are available for certain special cases even beyond two spatial dimensions. For example, we
have previously obtained the action for the edge modes
for quantum Hall droplets in $\mathbb{CP}^k$, both with constant and fluctuating Abelian and non-Abelian background gauge 
fields \cite{{KN2},{KN3},{Kar}}.
In section 4, we also carry out a direct quantization of these actions to obtain the current algebra. The results are shown to agree with the cocycle derived from the topological arguments.

Section 5 is devoted to a short discussion of the results
delineating the differences between the two-dimensional and higher dimensional situations. We give a summary of the
main results in section 6.
There are also four
short appendices outlining some technical results 
on deriving the bulk action from the Dolbeault index density, some useful
formulae, the edge actions and on
the canonical quantization of the edge modes.
\section{The two-cocycle and the commutator anomaly}
In this section we present the key arguments for relating the commutator anomaly to a two-cocycle. This will be done in two parts.
In the first part of this section we 
will consider the commutator anomaly for a field theory in its own right.
In the second part we will then consider this field theory as describing
the edge dynamics
of a theory in one higher dimension, thereby linking the commutator anomaly to a bulk action in the higher dimension.

\subsection{Commutator anomaly in field theory}

We start with a brief discussion of anomalies in terms of partition functions or
effective actions \cite{anomalies1}-\cite{anomalies3}. 
While the commutator anomaly is discussed in these papers, the argument we present here is somewhat independent and complete in its own right.

Consider
a field theory of scalars and/or fermions, generically denoted by
$\Phi$, and gauge fields, denoted by $A$,
in an even dimensional spacetime $\M_{2k}$, say of dimension $2k$.
The action is denoted by $S (\Phi, A)$ and the
partition function $Z(A)$ or effective action for the gauge fields
is given by the functional integral of $e^{i S}$ over the matter fields $\Phi$;
i.e.,
\beq
Z(A) = e^{i S_{\rm eff}(A)} = {\cal N} \int [D\Phi] \, e^{i S (\Phi, A)}
\label{anom1}
\eeq
where ${\cal N}$ is a normalization constant.
The symmetry transformation of interest will be of the form
$\Phi \rightarrow \Phi^g = g\, \Phi$ for matter fields
and $A \rightarrow A^g = g A g^{-1} - dg \, g^{-1}$ for gauge fields,
where $g \in G$. Generically, we take $G$ to be a Lie group
with elements of the form $g = e^{ i t_a \theta^a}$, $t_a$ forming a
hermitian basis of the generators of $G$.
The action $S(\Phi, A)$ has the symmetry
$S (\Phi^g, A^g ) = S(\Phi, A)$.
However, the regularization needed to define the functional integration
can render this symmetry anomalous.
This can be viewed as due to the nontrivial Jacobian resulting from the change of variables $\Phi \rightarrow \Phi^g$, i.e., from
\beq
[D \Phi^g] = e^{i \beta (g, A)} \, [D\Phi]
\label{anom2}
\eeq
Thus, correspondingly, we can write
\beqar
Z(A^g ) &=& {\cal N} \int [D \Phi] \,
e^{i S (\Phi, A^g)} \nonumber\\
&=& {\cal N} \int [D\Phi^g] \, e^{i S (\Phi^g, A^g)}
= {\cal N} \int [D\Phi] \, e^{i S (\Phi, A)}\, e^{i \beta (g, A)} \nonumber\\
&=& e^{i \beta (g, A)}  Z(A)
\label{anom3}
\eeqar
In going from the first line to the second we make a semantic change of
variables for the integration from $\Phi$ to $ \Phi^g$. Then since
$S (\Phi^g, A^g ) = S(\Phi, A)$, we can use (\ref{anom2}) to get to next
equality and the final result.
In (\ref{anom3}), $\beta (g, A)$ is in the form of an integral over $\M_{2k}$ of polynomials of the fields and their derivatives along with factors of $g$ and
$g^{-1} d g$, with the condition $\beta (1, A ) = 0$.
Typically, $\beta (g, A)$ also involves an integration over a line on the group
connecting the identity ($g =1$) to the finite element $g$.
The key point is that $\beta (g, A)$ cannot be written as the variation under
$g$ of any expression $S_c$ which is of the form of an integral of the fields and their derivatives. If this were possible, one could remove the anomaly
$\beta (g, A)$ by redefining the action as
$S \rightarrow S- S_c$. This would be equivalent to a different choice of regularization.

The group composition law imposes a consistency requirement on
$\beta (g, A)$. The action of $g g'\in G$ on the fields may be viewed
as the transformation by $g'$ followed by the action of $g$.
In other words $\Phi^{gg'}= g g' \Phi = (\Phi^{g'})^g$,
$A^{gg'} = (A^{g'})^{g}$. Applied to $Z(A)$, we find
\beqar
Z(A^{gg'}) &=& e^{i \beta (gg', A) } \, Z(A)\nonumber\\
&=&e^{i \beta (g, A^{g'}) } \, Z(A^{g'}) = 
e^{i [\beta (g, A^{g'})  + \beta (g', A)]}\, Z(A)
\label{anom4}
\eeqar
This shows that the anomaly $\beta (g, A)$ should obey the condition\footnote{Since we discuss the effective action here, we allow for
integrations-by-parts with vanishing boundary contributions and
the right hand side of (\ref{anom5}) is zero. More generally, as we shall see shortly, the right hand side will be the
integral of a total derivative.}
\beq
 \beta (gg', A)  - \beta (g, A^{g'})  - \beta (g', A) = 0
 \label{anom5}
 \eeq
 This is the so-called Wess-Zumino consistency condition
 \cite{WZ}.
 The anomaly $\beta (g, A)$ is defined on the space of fields, but it also has an integration along a line on the group from $g= 1$ to $g$, so it
 can be considered as a 1-cochain. The consistency requirement
 (\ref{anom5}) is then a closure condition on $\beta (g, A)$, so it becomes 
 a 1-cocycle.
 
So far the action and the effective action involve integration over all
of $\M_{2k}$. For defining the wave functions, we separate off the time-coordinate, taking spacetime to be of the form
$V_{2k-1} \times {\tau}$, where $V_{2k-1}$ denotes the spatial
manifold of dimension $(2k -1)$ and $\tau$ denotes
a segment of the real line $\mathbb{R}$, corresponding to time.
We will take $\tau = [ t_{\rm i}, t_{\rm f}]$, where $t_{\rm i}$ denotes the initial time
and $t_{\rm f} $ will be the final value of $t$ under consideration.
The wave function of the theory at time $t_{\rm f}$ is defined by
the functional integral
\beq
\Psi [ \Phi_{\rm f}, A_{\rm f}] = {\cal N} \,\int [D \Phi] d\mu (A) \, e^{i S (\Phi, A)}
\label{coc1}
\eeq
where we integrate over all fields obeying the boundary condition that
$\Phi (t_{\rm f}, \vx ) = \Phi_{\rm f} (\vx )$, $A(t_{\rm f}, \vx ) = A_{\rm f}(\vx)$,
and ${\cal N}$ is again some
normalization factor. The functional measure for the gauge fields
is the invariant measure on the space of gauge-orbits.
The specific state for which $\Psi [\Phi_{\rm f}, A_{\rm f}]$ is the wave function as given
in (\ref{coc1}) is determined by
the initial choice of the state (at time $t_{\rm i}$), which we do not display here explicitly
since it is not relevant to arguments which follow.

As before, the existence of an anomaly for the action of $G$ 
will involve the nontrivial transformation of the measure of functional
integration. We then get
\beq
\Psi (\Phi_{\rm f}^g, A_{\rm f}^g ) = \int [D\Phi^g]\,d\mu(A^g) \, e^{i S (\Phi^g, A^g)}
= \int [D\Phi]\,d\mu(A)\, e^{i S (\Phi, A)}\, e^{i \beta (g, A)} 
\label{coc2}
\eeq
The group composition law for the transformations now
gives
\beqar
\Psi \bigl( (\Phi_{\rm f}^{g'})^g,
(A_{\rm f}^{g'})^g\bigr)&=& \int [D\Phi^{g'}]\,d\mu(A^{g'})\, e^{i S (\Phi^{g'}, A^{g'})}\, e^{i \beta (g, A^{g'})} \nonumber\\
&=& \int [D\Phi]\, d\mu(A)\,e^{i S (\Phi, A)}\, e^{i \beta (g, A^{g'})} 
e^{i \beta (g', A)} \label{coc3}\\
\Psi \bigl(\Phi_{\rm f}^{gg'}, A_{\rm f}^{gg'} \bigr) &=& \int [D\Phi]\,d\mu(A)\, e^{i S (\Phi, A)}\, e^{i \beta (gg', A)}
\label{coc4}
\eeqar
These two results should match since
$(\Phi_{\rm f}^{g'})^g = \Phi_{\rm f}^{gg'}$, 
$(A_{\rm f}^{g'})^g = A_{\rm f}^{gg'} $.
Generally, and as we shall see explicitly a little later,
the difference of $\beta$'s in (\ref{coc3}) and (\ref{coc4}) is of the form
\beq
 \beta (gg', A)  - \beta (g, A^{g'})  - \beta (g', A) = \int d\omega(g, g', A)
 \label{coc5}
 \eeq
where $\omega(g, g', A)$ is a 2-cochain. In the case of integration over all time
with appropriate vanishing conditions on the boundary of
$\M_{2k}$, the right hand side integrates to zero in agreement
with (\ref{anom5}) for the effective action.
However, in the present case, we are integrating over
$V_{2k-1} \times {\tau}$, so the right hand side of
(\ref{coc5}) leads to integrals on the boundary.
We will choose the fields to vanish
on the spatial boundary, so that, effectively, we can take
$\del V_{2k-1} = \emptyset$, as far as the fields of interest are concerned.
On the right hand side of (\ref{coc5}), we are then left with $\oint_{V_{2k-1}} \omega(g, g', A_{\rm f})$ at $t_{\rm f}$
and a similar term at $t_{\rm i}$.
The integral at $t_{\rm f}$ can be taken out of the functional integration since it involves only data at $t_{\rm f}$ and
we can write (\ref{coc4}) as
\beqar
\Psi \bigl(\Phi_{\rm f}^{gg'}, A_{\rm f}^{gg'} \bigr) &=& \int [D\Phi]\,d\mu(A)\, e^{i S (\Phi, A)}\, e^{i \beta (gg', A)}\nonumber\\
&=& e^{i \oint_{V_{2k-1}} \omega(g, g', A_{\rm f})} \, 
\int [D\Phi]\, d\mu(A)\,e^{i S (\Phi, A)}\, e^{i \beta (g, A^{g'})} 
e^{i \beta (g', A)}\nonumber\\
&=&e^{i \oint_{V_{2k-1}} \omega(g, g', A_{\rm f})} \,  
\Psi \bigl( (\Phi_{\rm f}^{g'})^g,
(A_{\rm f}^{g'})^g\bigr)
\label{coc6}
\eeqar
If we consider the gauge transformation on the wave function to be implemented by the action of an operator $U(g)$, we can write
\beq
\Psi (\Phi_{\rm f}^g, A_{\rm f}^g) = U(g) \, \Psi (\Phi_{\rm f}, A_{\rm f})
\label{coc7}
\eeq
The relation (\ref{coc6}) then implies that the composition law is realized at the level of the operators $U(g)$ as
\beq
U(g g') = e^{i \oint_{V_{2k-1}} \omega(g, g', A_{\rm f})} \,  U(g) \, U(g') 
\label{coc8}
\eeq
This shows that we have a projective realization of the action of
$U(g)$ on the wave functions. At the level of infinitesimal generators,
this rule implies that we have a commutator anomaly.
Let $\theta$ and $\theta'$ denote the group parameters
corresponding to $g$ and $g'$, respectively, with $g = e^{i t_a \theta^a}$,
$g' = e^{i t_a \theta'^a}$.
We then have, for small values of the parameters,
\beq
g g' = e^{ i t_a \left(\theta^a + \theta'^a - {1\over 2} (\theta\times \theta')^a
+ \cdots \right)}, \hskip .3in (\theta\times \theta')^a
= f^{abc} \theta^b \theta'^c
\label{coc9}
\eeq
Here $f^{abc}$ are the structure constants of the Lie algebra defined by
$[t^a, t^b ] = i f^{abc} t^c$.
Also let $G(\theta)$ denote the operator which generates the gauge
transformation, so that we can write
$U(g) = e^{i G(\theta)}$. $G(\theta)$ is linear in $\theta^a$,
so we get
\beqar
U(gg') &\approx& e^{i G(\theta + \theta' - {\half} \theta\times \theta' + \cdots)},
\nonumber\\
U(g) \, U(g') &=& e^{i G(\theta)} \, e^{i G(\theta')} 
\approx e^{i G(\theta) + i G(\theta') - {\half} [G(\theta), G(\theta')] +\cdots}
\label{coc10}
\eeqar 
It is now easy to see that (\ref{coc8}) becomes
\beq
[G(\theta), G(\theta')] = i G( \theta\times \theta') + 2\, i \oint_{V_{2k-1}}
\omega (\theta, \theta', A_{\rm f})
\label{coc11}
\eeq
This displays $\oint \omega$ as a commutator anomaly.
One can also consider
the triple combination $g_1 g_2 g_3$. The reduction of this,
in consistency with associativity of group transformations, leads to
a condition on $\omega(g, g', A_{\rm f})$. This is a closure condition,
qualifying $\int \omega (g, g', A_{\rm f})$ as a 2-cocycle.
A similar result holds at the infinitesimal level for 
$\omega (\theta, \theta', A_{\rm f})$
in (\ref{coc11}); it is equivalent to the Jacobi identity.

It is also useful to relate $\omega (\theta, \theta', A_{\rm f})$ to the descent equations
for anomalies. (See lectures by Jackiw and by Zumino in \cite{anomalies2}.)
Let $\omega_{2 k+1}$ denote the Chern-Simons term associated to an index density $\I_{2k+2}$ in $2k+2$ dimensions; i.e.,
$d \omega_{2 k+1} = 2\pi \,\I_{2k +2}$.\footnote{The factor of $2\pi$ ensures that we have the correct normalization for the anomaly.}
$\omega_{2k +1}$ is a $(2k+1)$-form
made of gauge fields, spin connections, curvatures, etc.
The gauge fields are connections on a principal fiber bundle
and gauge transformations can be represented by exterior differentiation,
denoted by $\delta$, along the fiber directions. Writing the bundle connection as $\A = g A g^{-1} - d g \, g^{-1}$, we see that
\beqar
d\,\A + \A \, \A &=& g\, F\, g^{-1}\nonumber\\
\delta \A &=& d v + \A v + v \A = \D v , \hskip .2in v = \delta g\, g^{-1},
\label{coc12}
\eeqar
keeping in mind that $d\delta + 
\delta d = 0$.
The gauge invariance of $\I_{2k +2}$ tells us that 
\beq
d (\delta \omega_{2k +1}) = - \delta (d \omega_{2k+1})
= - 2\pi ~\delta \I_{2k+2} 
= 0
\label{coc13}
\eeq
so we have a locally defined form $\alpha^{(1)}_{2 k}$
such that $\delta \omega_{2k+1} = d \alpha^{(1)}_{2 k}$.
The superscript on $\alpha^{(1)}_{2 k}$ indicates that it has one power of
$v = \delta g \, g^{-1}$.
Interpreting $v $ as $\theta$ (for small $\theta$), $\int (-\alpha^{(1)}_{2 k})$
is the anomaly for the $2k$-dimensional field theory; i.e.,
\beq
\beta (g , A)\Big\vert_{g \approx 1 + i t_a \theta^a}
\approx  -  \int_{\M_{2k}} \alpha^{(1)}_{2 k} (\theta, A) \hskip .2in
\label{coc14}
\eeq

Since $\delta^2 = 0$, we also have $d (\delta \alpha^{(1)}_{2 k}) = 0$,
implying that we can write $\delta \alpha^{(1)}_{2 k} = d \omega^{(2)}_{2k -1}$.
The differential form $\omega^{(2)}_{2k -1}$ will have two powers of $v$, antisymmetrized,
which may be identified as $\theta$ and $\theta'$.
The integral of $\omega^{(2)}_{2k -1}$ over the $(2k-1)$-dimensional spatial manifold is the commutator anomaly, i.e.,
\beq
\oint_{V_{2k-1}} \omega(\theta, \theta', A_{\rm f}) = 
- \oint_{V_{2k-1}} \omega^{(2)}_{2k -1} (\theta, \theta', A_{\rm f})
\label{coc15}
\eeq
Reasoning along similar lines as before, $\delta \omega^{(2)}_{2 k -1} =
d \omega^{(3)}_{2k -2}$, so that $\oint \omega^{(2)}_{2 k -1}$ is $\delta$-closed,
i.e., $\delta \oint \omega^{(2)}_{2 k -1} = 0$. This implies that
$\oint \omega^{(2)}_{2 k -1}$ is a 2-cocycle. As mentioned above, at
the level of commutators, this is equivalent to the Jacobi identity.

Although we outlined the descent procedure using the exterior derivative
$\delta$, one can also identify $\omega$ by carrying out infinitesimal
transformations and antisymmetrizing.
Since $\alpha^{(1)}_{2k}$ has one power of the (infinitesimal)
parameter $\theta$, we may write (\ref{coc14}) as $\beta\approx -\int \alpha (\theta)$. The condition (\ref{coc5}) then gives
\beq
\delta_\theta \, \alpha (\theta', A) - \delta_{\theta'} \, \alpha (\theta, A)
- \alpha (\theta \times \theta', A) = - 2 \oint_{V_{2k-1}} \omega (\theta, \theta', A_{\rm f})
\label{coc16}
\eeq
This gives a straightforward way to identify $\omega$.

In this section we have given a general argument on how to relate the
commutator anomaly in a field theory to the 2-cocycle derived from 
Chern-Simons term via the descent procedure.

\subsection{Bulk and boundary theories}

So far we have considered a field theory in $2k$-dimensional spacetime in its own right. 
We now consider the placement of the argument given above in the context of a higher dimensional field theory.
Towards this, consider
a $(2k+1)$-dimensional spacetime manifold
of the form $\M_{2k+1} = V_{2k} \times \tau$, where the boundary of
$V_{2k}$ is the space $V_{2k-1}$ considered above.
The bulk action for $\M_{2k+1}$ should be such that its variation under the symmetry transformation produces a boundary term which is canceled
by the anomaly of the boundary theory.
In other words, there is a cancellation of the anomaly between the bulk and boundary dynamics and the
full theory preserves the gauge symmetry.
All the considerations of anomalies in this section up to the previous
paragraph are to be taken as applying to the boundary theory
on $V_{2k-1} \times \tau$.

Since the anomaly of the boundary theory is $-\int \alpha^{(1)}_{2k}$,
it is easy to see that the action for the $(2k+1)$-dimensional theory
should be
\beq
S =  \int_{\M_{2k+1}}\!\! \!  \omega_{2k+1}  ~+ S( \Phi, A )
\label{coc17}
\eeq
where $S(\Phi, A)$ is the action for the theory
considered in (\ref{coc1}), which is defined on the manifold
$V_{2k-1 }\times \tau$. The action (\ref{coc17}) is free of anomalies,
the anomaly of the boundary theory canceling against the
variation of the bulk part of the action.

The cancellation between the gauge variation of the bulk action and the anomaly of the boundary degrees of freedom is also known as the anomaly-inflow mechanism, familiar in particular from the work of Callan and Harvey \cite{CallanHarvey}. In the present application, the bulk quantum Hall Chern-Simons action supplies the inflow, while the chiral edge theory carries the corresponding boundary anomaly. Our use of this mechanism is to infer the boundary current-algebra two-cocycle from the known bulk action, even when an explicit edge action is unavailable.

We are going to apply these ideas to the case of the quantum 
Hall effect. The effective bulk actions for QHE with integer filling fraction $\nu$ on
complex manifolds of arbitrary even
dimensions were obtained in \cite{KN-dolb} using the 
Dolbeault index theorem. The action is in fact the Chern-Simons term
$\omega_{2k+1}$ corresponding to the Dolbeault index density
$\I_{2k +2}$.
The background values for the gauge and gravitational fields in these actions were arbitrary, except for being constrained to a particular topological class.

We also note that the bulk action for the Laughlin-type
states with fractional values of $\nu$ with Abelian gauge fields were obtained for (2+1) and (4+1) dimensions
in \cite{AKN}, as a generalization of the parton picture
previously used in two dimensions.
The main observation here is that the bulk action 
is easier to construct and is known for general backgrounds, 
as we indicate in Appendix A.

By contrast, the situation for the boundary action is more limited.
For the case of the integer QHE on $\mathbb{CP}^k$ manifolds,
with Abelian and non-Abelian background magnetic fields proportional to the ``constant" $U(k)$ curvatures of $\mathbb{CP}^k$,
the boundary action was obtained in
\cite{{KN2},{KN3}} using the coadjoint orbit method.
The result is basically a higher dimensional generalization of the 2d chiral bosonic action. Fluctuating gauge fields were also introduced via a gauged version of the coadjoint orbit method in \cite{{Kar},{reviewKN}},
leading to an effective action with
cancellation of anomalies between
the bulk and the boundary.
All these results were for the fixed geometric background of $\mathbb{CP}^k$
and in a large-$n$ limit.

We can now see that the arguments presented here
lead to a simple and general strategy for obtaining the commutator anomaly
or the nontrivial extension of the current algebra associated to 
the gauge transformations of interest for the chiral edge modes
of the quantum Hall droplet.
As mentioned above, the bulk actions are known for general backgrounds.
So we can obtain the corresponding two-cocycle via the descent procedure
mentioned above. The arguments given above then show that
the two-cocycle given in (\ref{coc11}), and which is constructed via descent from
the Chern-Simons form as in (\ref{coc15}) or (\ref{coc16}),
can be identified as the extension term for the current
algebra acting on the states of the boundary theory.

The key point is that
even if we do not have the general action for the chiral edge modes, we can determine the corresponding current algebra for the edge 
modes by identifying the 
two-cocycle from the bulk action, the latter, as mentioned above, being known 
in full generality.
It is the cancellation between the gauge variation of the bulk action and
the anomaly of the boundary theory which makes this possible.
This is the main point of this paper. 
We will now work out some specific cases in the next section.
We will also show that, in cases where the edge action is known,
such as the case of QHE on $\mathbb{CP}^k$ as mentioned above,
the extended current algebra obtained by these topological arguments also agrees with the direct quantization of the edge theory.
\section{Bulk actions from an index theorem}
In the Appendix A, we have briefly recapitulated the arguments
based on the Dolbeault index theorem \cite{eguchi} from which one can 
obtain the form of the bulk effective action for a quantum Hall
droplet of fermions \cite{KN-dolb}. For simplicity we will only consider
the case of the droplets in the lowest Landau level.
In a general dimension, the bulk action for $\nu =1$  takes the form
\begin{equation}
S^{\rm bulk} =
 \int \Bigl[ {\rm td}(T_c K) \wedge \sum_p  (CS)_{2 p+1} ( A)\Bigr]_{2 k+1}
+ S^{\rm grav}
\label{bulk1}
\end{equation}
The Chern-Simons term
$(CS)_{2 p +1}(A)$ is defined by
\begin{equation}
{1\over 2 \pi} d (CS)_{2 p +1}(A) = {1 \over (p+1)!} \Tr \left(
{ i F \over 2 \pi}\right)^{p +1}
\label{bulk2}
\end{equation}
and  ${\rm td}(T_c K)$
is the Todd class for the complex tangent space, see Appendix A for more details.
(The Chern-Simons term is the same as the $\omega_{2p+1}$
defined earlier if the index density is just the Chern character, i.e., the purely $A$-dependent right hand side of ({\ref{bulk2}).)
In the expansion of the Todd class and the $CS$ terms
in (\ref{bulk1}), we should use the differential form of rank $2k+1$
for the action, as indicated by the subscript $2k+1$.

The purely gravitational part of the action $S^{\rm grav}$ is
defined by
\beqar
S^{\rm grav} &=& 2 \pi\int \Omega^{\rm grav}_{2k+1}\nonumber\\
d\, \Omega_{2 k+1}^{\rm grav} 
&=& \left[ {\rm td}(T_cK) \right]_{2 k +2}
\label{dolb11}
\eeqar

The expression (\ref{bulk1}) is general, but it is useful to
write it out explicitly for some low dimensions.
For this, we first note that the $CS$ terms are given by
\beqar
(CS)_1 &=& i \Tr (A)\nonumber\\
(CS)_3 &=&{ i^2 \over 4\pi} \Tr \left( A F - {1\over 3} A^3 \right)\nonumber\\
(CS)_5&=& {i^3 \over 3! (2 \pi)^2} \Tr \left( AF^2 - {1\over 2} A^3 F 
+ {1\over 10} A^5\right) \nonumber\\
(CS)_7&=&{1\over 4! (2 \pi)^3} \Tr
\left( AF^3 - {2\over 5}A^3 F^2 - {1\over 5} (A F A^2 F - A^4 F) - {1\over 35} A^7
\right)
\label{dolb11a}
\eeqar
The traces involved in the Todd class will depend on the dimension
since the curvature of the manifold takes values in
the algebra of the holonomy group $U(k)$.
In 2+1 dimensions, for which $k =1$, so $U(k) = U(1)$, we get
the bulk action
\begin{equation}
S^{\rm bulk}_{2+1} 
= {i^2 \over {4\pi}} \int \Biggl\{ A dA + A d \omega + {1\over 6} \omega d\omega  \Biggr\}
\label{dolb12}
\end{equation}
where $\omega$ is the spin connection.
The fields $(A, \omega)$ are given in
an antihermitian basis to be uniform in notation with
what we used in earlier sections and in (\ref{dolb11a}).
(Thus $A$ in this equation stands for
$- i $ times the real gauge field.)
 The last term, which is independent of the gauge fields will not be important for the commutator anomaly for the gauge transformations.

In 4+1 dimensions, the bulk action becomes
\begin{eqnarray}
S^{\rm bulk}_{4+1} &=& \int \left[ (CS)_{5} + {c_1 \over 2} \, (CS)_3 + {c_1^2 + c_2 \over 12} (CS)_1\right] + S^{\rm grav}\nonumber\\
&=&i^3\int  \Biggl\{ \left[ {1 \over 3! (2 \pi)^2} \Tr \left( AF^2 - {1\over 2} A^3 F 
+ {1\over 10} A^5\right) \right]
+ {d \omega^0 \over 8 \pi^2} \Tr\left( A F - {1\over 3} A^3\right)\nonumber\\
&&\hskip .4in + \left[ {5 \over 48 \pi^2} d\omega^0\, d\omega^0
- {1\over 96 \pi^2} \Tr ( {\tilde R} {\tilde R} ) \right] \Tr A 
\Biggr\} + S^{\rm grav}
\label{dolb13}
\eeqar
Again, we use an antihermitian basis. In 4+1 dimensions, the holonomy group is $U(2)$ since $k =2$. Thus the curvature is of the form
\beq
R = d \omega^0 \mathbb{1} + (-i t_a ) R^a
= d \omega^0 \mathbb{1} + {\tilde R}
\label{dolb3a}
\eeq
Here $\mathbb{1}$ is the $2\times 2$ identity matrix and 
$t_a$ are the generators of $SU(2)$ in the fundamental
representation, with $\Tr (t_a t_b ) = {\half} \delta_{ab}$.
We have not displayed the purely gravitational part since it is not relevant
for the commutator anomaly.

In 6+1 dimensions, we can similarly work out the bulk action as
\beqar
S^{\rm bulk}_{6+1}&=& \int \left[ (CS)_7 + {c_1 \over 2} (CS)_5 + {c_1^2 + c_2 \over 12}
(CS)_3 + {c_1 c_2 \over 24} (CS)_1\right] + S^{\rm grav}\nonumber\\
&=&\int \Biggl\{ (CS)_7 + i {3  \over 4\pi} d\omega^0 \,(CS)_5
+ {i^2 } \left[ {1 \over 4 \pi^2} (d \omega^0)^2 - {1\over 96\pi^2} \Tr ({\tilde R} {\tilde R} ) \right] (CS)_3 \nonumber\\
&& \hskip .3in
+ i^3 \left[ {3 \over 64 \pi^3} (d \omega^0)^3 - {d\omega^0 \over 128 \pi^3}
\Tr ({\tilde R} {\tilde R} ) \right] (CS)_1 \Biggr\} + S^{\rm grav}
\label{dolb14}
\eeqar
The curvature has the form given in (\ref{dolb3a}), except that
$\mathbb{1}$ and $t_a$ are now $3\times 3$ matrices
since the holonomy group is $U(3)$.

In all the cases of QHE on arbitrary even spatial dimensions analyzed in \cite{KN1}-\cite{KN3}} an Abelian background field
of the form $d A = -i  n \Omega_{\rm K}, $\footnote{$\Omega$ is written in antihermitian form in accordance with the convention we have used. 
However, we use hermitian $\Omega_{\rm K}$ since that is more conventional
in the literature on complex projective spaces.} where $\Omega_{\rm K}$ is the
K\"ahler two-form on the spatial manifold, was used, in addition to any non-Abelian background fields or fluctuations of the gauge potentials.
A high magnetic field (obtainable for large $n$) is needed to have a large number of states and large occupancy so that one can have a well-defined droplet of fermions. The actions for the
edge modes of the droplet were also worked out.
It is therefore useful to expand the bulk actions given above around the background value of $A$
by shifting $A \rightarrow A+ a$, where 
$d a = \Omega = -i  n \Omega_{\rm K}$.
This expansion is important in obtaining the two-cocycle in a form which can then be compared to the direct quantization of the edge actions in
\cite{{KN2},{KN3},{Kar}}.
It is easy to check that\footnote{For the bulk actions, to identify the correct local action densities in terms of $\Omega$, certain integrations-by-parts 
are allowed and have been carried out.}
\beq
(CS)_{2 p+1} (A + a) = \left[ \left( e^{ i d a/2\pi} \right) (CS)(A) \right]_{2p+1}
+ (CS)_{2p +1} (a)
\label{dolb14a}
\eeq
Keeping only the $A$-dependent terms, where $A$ now is of order $n^0$, we find,
\beqar
S^{\rm bulk}_{2+1}&=& {i^2 \over 4\pi} \int \left[ A dA + A d\omega + 2\, \Omega A \right]\nonumber\\
S^{\rm bulk}_{4+1}&=&\int \Biggl\{ (CS)_5 + { i (\Omega + d\omega^0)\over 2\pi} (CS)_3
\nonumber\\
&&\hskip .3in+ {i^2 \over 8 \pi^2}\left[ \Omega^2 + 2  \Omega \, d\omega^0 + {5 \over 6}
d\omega^0 d\omega^0 - {1\over 12} \Tr ({\tilde R} {\tilde R})
\right] (CS)_1\Biggr\}
\label{dolb14b}\\
S^{\rm bulk}_{6+1}&=&\int \Biggl\{ (CS)_7 + i \left( {\Omega\over 2\pi}  +  {3 d\omega^0\over 4\pi}\right) (CS)_5
\nonumber\\
&&\hskip .3in+ {i^2 \over 8 \pi^2}\left[ \Omega^2  + 3  \Omega\, d\omega^0 + 2 d\omega^0 d\omega^0  - {1\over 12} \Tr ({\tilde R} {\tilde R})
\right] (CS)_3\nonumber\\
&&\hskip .3in +{i^3 \over 3! (2\pi)^3} \biggl[ \Omega^3 + {9 \over 2} \Omega^2  d\omega^0
+ 6 \Omega (d\omega^0)^2 + {9 \over 4} (d \omega^0)^3\nonumber\\
&&\hskip 1in - 
\left( {\Omega \over 4} + {3 d\omega^0 \over 8}\right) \Tr ({\tilde R} {\tilde R})
\biggr] (CS)_1\Biggr\} 
\label{dolb14c}
\eeqar

A similar analysis was carried out for some fractional values of the 
filling fraction $\nu$ in \cite{AKN}. The strategy was to use 
the so-called parton picture
where the fundamental charge carrier, the electron, is viewed as
made of several partons of fractional charge.
Additional gauge fields are needed to obtain the binding of the partons
to form the electron. Equivalently, these fields lead to constraints
which ensure that the observable charge carriers are the electrons.
{\it A priori}, there are additional anomalies due to the gauge symmetries of these
additional gauge fields. The cancellation of these anomalies, needed to ensure that the constraints can be consistently implemented, then lead to the
fractional states. The resulting bulk actions were worked out in
\cite{AKN} for 2+1 and 4+1 dimensions for Abelian gauge fields. For filling fraction
$\nu = 1/m$, these are given by
\beqar
S^{\rm bulk}_{\rm 2+1, \,\nu = 1/m}
&=& {i^2 \over 4 \pi} \int \left[ { A dA \over m} + A \, d\omega 
\right] +
 {i^2 \over 4 \pi} \int \left( {m \over 4} - {1\over 12} \right) \omega d\omega
 \nonumber\\
 &\xrightarrow{A \rightarrow A + a} & {i^2 \over 4 \pi} \int \left[ { A dA \over m} + A \, d\omega  
 + {2 \Omega A \over m}
\right] + \cdots
\label{dolb15}
\eeqar
The result for 4+1 dimensions is
\beqar
S^{\rm bulk}_{\rm 4+1, \,\nu = 1/m}&\!=\!\!&i^3 \int \Biggl\{ {1\over 24\pi^2} {A F^2 \over m^2}
+ {1 \over 8 \pi^2} d \omega^0 {AF \over m} 
+ \left[ {5 \over 48 \pi^2} (d \omega^0)^2 - {1\over 96 \pi^2} \Tr ({\tilde R}
{\tilde R}) \right] A 
 \Biggr\}\nonumber\\
 &&\hskip .2in  + S^{\rm grav}\nonumber\\
 &\xrightarrow{A \rightarrow A +a}&i^3 \int \Biggl\{ 
 {1\over 24\pi^2} {A F^2 \over m^2} + {1\over 8 \pi^2} \left( {\Omega \over m^2}
 + {d \omega^0 \over m} \right) A F
\nonumber\\
&&\hskip .3in+{1 \over 8 \pi^2} \left( {\Omega^2 \over m^2} + 2 {\Omega\, d\omega^0 \over m}
+ {5 \over 6} (d\omega^0)^2 - {1\over 12} \Tr ({\tilde R} {\tilde R})
\right) A \Biggr\}+ \cdots
\label{dolb16}
\eeqar
The anomaly cancellation for the partons used in \cite{AKN} to 
formulate the fractional Hall states is significantly more involved
for non-Abelian background gauge fields, so, in (\ref{dolb15}), (\ref{dolb16}),
we have given the results only for Abelian gauge fields.
\section{The commutator anomaly for the quantum Hall droplets}
The commutator anomaly can now be derived in a straightforward way.
We are interested in gauge transformations of the gauge field
$A$. Therefore, for now, the $A$-independent terms, i.e., the purely gravitational terms in the action, are not of interest.
Further, since the commutator anomaly involves two gauge transformations (with parameters $\theta$, $\theta'$), we need at least two factors of
$A$ to get a nonzero contribution. So the terms involving $(CS)_1$ are not
important for this purpose.
From the gauge variation of the $CS$ terms we then
find
 \begin{align}
\int_{\M_{2k +1}} \delta_\theta (CS)_{2k+1}&= \int_{\M_{2k}} \alpha_{2k} (\theta, A)\nonumber\\
\alpha_2 (\theta, A) &= {i^2 \over 4\pi} \Tr [ d \theta \, A] \nonumber\\
\alpha_4 (\theta, A) &= {i^3 \over 24\pi^2} \Tr \left[ d \theta \, \left(A dA + {1\over 2} A^3\right)\right] \nonumber\\
&={i^3 \over 24\pi^2} \Tr \left[ d \theta \, \left(A F - {1\over 2} A^3\right)\right] \label{comm1}\\
\alpha_6 (\theta, A) &= {i^4 \over 24(2\pi)^3} \Tr \left[ d \theta \, \left(A dA dA  - {1\over 5} (A^2 d A\, A - 3 A dA\,A^2 - 4 A^3 dA)+ {2\over 5} A^5\right)\right] \nonumber\\
&={i^4 \over 24(2\pi)^3} \Tr \left[ d \theta \, \left( A F^2 - {1\over 5} A^3 F
- {2\over 5} A F A^2 - {1\over 5} A^2 F A + {1\over 5} A^5\right)
\right]\nonumber
\end{align}
Carrying out a second variation of $\alpha_2$, we find 
\begin{align}
\delta_{\theta} \alpha_2 (\theta', A) - 
\delta_{\theta'} \alpha_2(\theta, A)&= 
- {i^2 \over 4\pi} \left[ \Tr (d \theta'\, d\theta
- d\theta d\theta')\right] - {i^2 \over 4\pi} \Tr \Bigl[( [\theta, d\theta'] +[d\theta, \theta'])A\Bigr]
\nonumber\\
&= - \alpha_2 ([\theta, \theta'], A) + {i^2\over 4\pi} \Tr ( d\theta d\theta'
- d\theta' d\theta)\nonumber\\
&= \alpha_2 (\theta\times\theta', A) + d\left[ {i^2\over 4\pi} \Tr (\theta d \theta' - \theta' d\theta)\right]
\label{comm2}
\end{align}
Comparing this with (\ref{coc16}), we see that the term in the
square brackets in this equation gives 
$- 2 \omega(\theta, \theta', A)$.
(The argument $A$ here is $A_{\rm f}$ if we compare this with
(\ref{coc16}).)
Similarly, carrying out the second variations we find
\beqar
\delta_\theta \alpha_4 (\theta', A) - \delta_{\theta'} \alpha_4 (\theta, A)
- \alpha_4 (\theta \times \theta', A) &=& d\left[ {i^3 \over 24 \pi^2} 
\Tr [(\theta d \theta' - \theta' d \theta) dA] \right]
\nonumber\\
\delta_\theta \alpha_6 (\theta', A) - \delta_{\theta'} \alpha_6 (\theta, A)
- \alpha_6 (\theta \times \theta', A) &=& d\left[
{1\over 24 (2\pi)^3} \Tr\left( ( d\theta d \theta' - d\theta' d\theta) \Bigl( A dA + {3\over 5}A^3 \Bigr)\right.\right.\nonumber\\
&&\hskip .3in\left. \left.+ {1\over 5} ( d \theta A d \theta' A^2 - d\theta' A d\theta A^2)\right) \right]
\label{comm2a}
\eeqar
We can thus identify the corresponding 2-cocycles as 
\beqar
2\, \omega_1(\theta, \theta' ,A)&=& - {i^2\over 4\pi} \Tr (\theta d \theta' - \theta' d\theta)\nonumber\\
2\, \omega_3(\theta, \theta' ,A)&=& - {i^3 \over 24 \pi^2} 
\Tr [(\theta d \theta' - \theta' d \theta) dA]
\label{comm2b}\\
2\, \omega_5(\theta, \theta' ,A )&=& -
{1\over 24 (2\pi)^3} \Tr\biggl[ ( d\theta d \theta' - d\theta' d\theta) \Bigl( A dA + {3\over 5}A^3 \Bigr)\nonumber\\
&&\hskip 1in + {1\over 5} ( d \theta A d \theta' A^2 - d\theta' A d\theta A^2)\biggr] \nonumber
\eeqar
where $\theta = i(\theta^0 + t_a \theta^a)$, and $\theta^0$, $\theta^a$ are the Abelian and non-Abelian real-valued transformation parameters. (For simplicity we have set $A_{\rm f} \equiv A$ in (\ref{comm2b}). 

Based on arguments given in section 2, the anomalous current algebra for the boundary theory defined on $\M_{2k}$ takes the form
\beq
[G(\theta), G(\theta')] = i G( \theta\times \theta') + 2\, i \oint_{V_{2k-1}}
\omega_{2k-1} (\theta, \theta', A)
\label{anoalg}
\eeq
where $G(\theta)$ is the generator of gauge transformations for the edge modes. 

In the next section we will give the anomalous current algebras for the edge modes in an explicit form for different dimensions {\it based solely on our knowledge of the bulk effective actions.} We will then
compare them with the current algebras obtained from the edge effective actions for QHE on $\mathbb{CP}^k$ we derived in \cite{{KN2}, {KN3}, {Kar}}. 

Before we proceed though, a comment about the given expressions for $\alpha_{2k}$ is appropriate.
Clearly the definition of $\alpha^{(1)}_{2k}$ via $\delta \omega_{2k+1}
= d \alpha^{(1)}_{2k}$ specifies the form of 
$\alpha^{(1)}_{2k}$ only up to a total derivative of the form
$d {\tilde \alpha}^{(1)}_{2k-1}$.  Being a total derivative, the contribution
of this term resides on the boundary and so
this may be viewed as an additional term in the definition 
of the generator of the gauge transformations for the edge modes,
i.e., $G(\theta) \rightarrow G(\theta) + {\tilde \alpha}^{(1)}_{2k-1}$.
Since $G(\theta)$ is in general a composite operator, 
regularization is needed and its renormalized form can have
a regularization dependence which is precisely of this form.
Thus in choosing a specific form for $\alpha^{(1)}_{2k}$, we are choosing a
particular regularization. This can change the form of $\omega_{2k-1}(\theta, \theta', A)$ in a cohomologically trivial way (see lectures by R. Jackiw in \cite{anomalies2}). 
Of course, the part of the commutator anomaly which is cohomologically significant is
unaltered by the ambiguity due to ${\tilde \alpha}^{(1)}_{2k-1}$.

\subsection{2+1 dimensions}
Comparison of these results with
(\ref{anoalg}) shows that the generator of
the gauge transformations on the wave functions of the
$1+1$ dimensional field theory (relevant to Hall droplets in
2+1 dimensions) should obey the anomalous algebra
\beq
[G(\theta), G(\theta')] = i G( \theta\times \theta') + { i \over 4\pi}\oint
\Tr (\theta d \theta' - \theta' d\theta)
\label{anoalg1}
\eeq
where $\theta = i (\theta^0 + t_a \theta^a)$.
The generator of gauge transformations on the matter fields on the boundary will correspond to the time-component of a suitable current, $J_0 (\theta) = \oint \bigl [\theta^0 J_0 (\vf) + \theta^a J^a_0 (\vf) \bigr]$, where $J_0 (\vf) , \, J^a_0 (\vf)$ are the Abelian and non-Abelian currents and
$\vf$ denotes the angular coordinate along the edge of the droplet. In this case, 
with $\Tr (t_a t_b ) = {\half} \delta_{ab}$, the algebra in (\ref{anoalg1}) reduces to
\beqar
[J_0 (\vf ), J_0 (\vf') ] & =& - {i \over 2\pi} \del_\vf \delta (\vf -\vf') \label{comm5} \\
{}[J_0^a(\vf ), J_0^b(\vf')] & = & i f^{abc} \, J_0^c (\vf) \, \delta (\vf-\vf')
- {i \over 4\pi} \delta^{ab} \del_\vf \delta (\vf-\vf')
\label{comm4}
\eeqar
Although here we display the non-Abelian current algebra as well
to show the nature of the central extension, in the
case of Hall effect in two spatial dimensions the gauge field of interest is Abelian, specifically, the electromagnetic field, and therefore only the algebra (\ref{comm5}) is relevant.

The result (\ref{comm5}) is obtained from the two-cocycle derived from the bulk
action. We can now compare with direct calculation from the edge dynamics.
It is known that the edge modes for the (1+1)-dimensional boundary theory
can be described by a chiral scalar field $\Phi$ \cite{Wen}, with action 
\beq
S = -{1\over 4\pi} \int dt d\vf\,\left[ {\dot \Phi} {\del \Phi \over \del \vf} + v_{\rm F} \left({\del \Phi \over \del \vf}
\right)^2 \right]
\label{comm6}
\eeq
where $v_{\rm F}$ is the Fermi velocity at the edge of the droplet.
(The action (\ref{comm6}) can also be derived using the coadjoint orbit
method, see Appendix C.)
The canonical quantization of this theory leads to the commutation rule
\beq
[\Phi(\vf), \Phi(\vf')] = i \pi \,[\Theta (\vf-\vf') - \Theta (\vf'-\vf)]
\label{comm7}
\eeq
where $\Theta (\vf - \vf')$ is the step function.
These commutation rules also show that
\beqar
[ -(1/2\pi) \del_\vf \Phi (\vf), \Phi (\vf') ] &=& -i\, \delta (\vf-\vf')
\nonumber\\
{}[ -(1/2\pi) \del_\vf \Phi (\vf), -(1/2\pi) \del_{\vf'} \Phi (\vf')]
&=& - {i\over 2\pi} \del_\vf \delta (\vf-\vf')
\label{comm8}
\eeqar
We see that we can therefore identify the canonical
generator of the gauge transformations as
\beq
J_0 (\theta^0 ) =  - {1\over 2\pi} \int d\vf~ \theta^0\, \del_\vf \Phi
\label{comm9}
\eeq
It is easily verified that the gauge transformation of $\Phi$ is
obtained as
$e^{i J_0 (\theta^0)} \Phi e^{- i J_0 (\theta^0)} = 
\Phi + \theta^0$. The second relation
in (\ref{comm8}) shows that the canonical quantization of the boundary theory
reproduces the commutator anomaly in (\ref{comm5}).

Turning to the fractional Hall effect of $\nu = 1/m$ with the action given in
(\ref{dolb15}), we see that we can easily modify our 
considerations above to get the anomalous algebra
\beq
[ G(\theta^0), G(\theta^{0\prime}) ] \equiv [J_0(\theta^0), J_0(\theta^{\prime 0})]  = - {i \over 2 \pi m} \oint \theta^0 \,\del \theta^{\prime 0}
\label{comm9a}
\eeq
This corresponds to an edge action of the form
\beq
S = -{1\over 4\pi m} \int dt d\vf\,\left[ {\dot \Phi} {\del \Phi \over \del \vf} + v_{\rm F} \left({\del \Phi \over \del \vf}
\right)^2 \right]
\label{comm9b}
\eeq

\subsection{4+1 dimensions}
For the (3+1)-dimensional boundary theory corresponding to
a droplet in quantum Hall effect in 4+1 dimensions, 
the generators of the gauge transformations obey the
anomalous algebra
\beq
[G(\theta), G(\theta')] - i G( \theta\times \theta') =
- {1 \over 8 \pi^2} \oint  \Tr \left[
(\theta d\theta' - \theta' d\theta)\left( {d A \over 3} + \Omega + d\omega^0
\right)\right]
\label{comm10}
\eeq
This result follows from (\ref{dolb14b}) and (\ref{comm2b}), i.e., just from the bulk action.
This is the general result for a complex spatial manifold, and we have kept terms of all orders in $n$ for the anomalous term on the right hand side.

In the case of the spatial manifold being $\mathbb{CP}^k$,
there is a direct calculation of the edge action at large $n$
using the method of coadjoint orbits \cite{{KN2}, {KN3}},
as summarized in Appendix C.
So we will now consider the simplification of the anomaly term
in (\ref{comm10}) for
$\mathbb{CP}^2$ since that will provide
a point of comparison with 
the quantization of the edge modes.
In general, for $\mathbb{CP}^k$, the
term proportional to $\Omega$ on the right hand side of 
(\ref{comm10}) will be the leading term at large $n$, since
$\Omega = -i n\, \Omega_{\rm K}$.
Here 
$\Omega_{\rm K}$ is the Fubini-Study K\"ahler form
\beq
\Omega_{\rm K} = i \left[ {dz^i \, d\bz^i \over (1+ \bz \cdot z)}
- { \bz\cdot dz\, z\cdot d\bz \over (1+ \bz \cdot z)^2}
\right] 
\label{comm11}
\eeq
The complex coordinates of $\mathbb{CP}^2$ are
$z^i$, $i = 1, 2$, which can be related to real local coordinates by
$z^1 = x^1+i x^2$, $z^2= x^3 + i x^4$.
Elementary algebraic relations from Appendix B  then lead to the result
\beq
- {1 \over 8 \pi^2} \oint  \Tr \left[
(\theta d\theta' - \theta' d\theta)\, \Omega \right]
= {i n \over 4 \pi^2} \oint dS_3\, {r_{\rm D}^2 \over (1+ r_{\rm D}^2)^3}
\Tr ( \theta \L \theta' - \theta' \L \theta )\label{comm13}
\eeq
where $\L$ denotes the angular derivative
\beq
\L =- 2\, r_{\rm D}\, {\hat x}_{\mu} (\Omega_{\rm K}^{-1})^{\mu\nu}
\del_{\nu} = 
 i \left( z\cdot {\del \over \del z} - \bz \cdot {\del \over \del \bz}
\right) \equiv  {\del \over \del \vf}
\label{comm13a}
\eeq
where $\vf$ is the common phase for all $z^i$.
We have considered a spherical droplet of radius $r_{\rm D}$
for simplicity. The integration in (\ref{comm13}), indicated by
$dS_3$, is over the three-sphere
which is the boundary of the droplet. Further, as is clear from
(\ref{comm13a}), $\L$ is the angular derivative on the surface along the direction Poisson-conjugate to the radial direction ${\hat x}_{\mu}$.

To facilitate comparison with the edge action derived in
\cite{{KN2},{KN3}}, and summarized in Appendix C,
we note that the
states of the lowest Landau level
for $\mathbb{CP}^2$ form a symmetric representation
of $SU(3)$ of rank $n$, which may be decomposed into a series of
$SU(2)$ representations. 
In \cite{{KN2},{KN3}}, we considered a droplet 
corresponding to the first $M$ $SU(2)$ representations to be
occupied. In this case $r_{\rm D}^2 = M/n$ for large $n$, $M$, with
$n \gg M$. We can then simplify the right hand side of (\ref{comm13})
to get
\beq
- {1 \over 8 \pi^2} \oint  \Tr \left[
(\theta d\theta' - \theta' d\theta)\, \Omega \right]
=
- i { M \over 2 \pi^2} \oint dS_3\, (\theta^0 \L \theta^{0\prime})
\label{comm14}
\eeq
This is for the Abelian case. In the non-Abelian case with
$\theta = i t_a\theta^a, \theta' = i t_a \theta'^a$, we get
\beqar
- {1 \over 8 \pi^2} \oint  \Tr \left[
(\theta d\theta' - \theta' d\theta)\, \Omega \right]
&=& i { M \over 2 \pi^2} \oint dS_3\,\Tr (\theta \L \theta')\nonumber\\
&=&- i { M \over 4 \pi^2} \oint dS_3\, (\theta^a \L \theta'^a)
\label{comm15}
\eeqar
The remaining terms in (\ref{comm10}) are subleading in 
$1/n$. Their simplification is more straightforward and gives
\beqar
&&- {1 \over 8 \pi^2} \oint  \Tr \left[
(\theta d\theta' - \theta' d\theta)\left( {d A \over 3} + d\omega^0
\right)\right]\nonumber\\
&&\hskip .5in 
= - {1 \over 16 \pi^2} \oint dS^{\nu\alpha\beta}
\Tr \left[(\theta \del_\nu \theta' - \theta' \del_\nu \theta) \left( {1\over 3} f_{\alpha\beta} + R^0_{\alpha\beta}\right)\right]
\label{comm16}
\eeqar
where 
\beq
dS^{\nu\alpha\beta} = r_{\rm D}^3
 \,{\hat x}_\mu \epsilon^{\mu\nu\alpha\beta} \, dS_3
 \label{comm16a}
 \eeq
Also
$f_{\alpha\beta} = \del_\alpha A_\beta - \del_\beta A_\alpha$ and
$R^0_{\alpha\beta} = \del_\alpha \omega_\beta^0 - \del_\beta \omega^0_\alpha$ is the $U(1)$ component of the curvature.

While all these terms (\ref{comm14})-(\ref{comm16}) can contribute to the commutator anomaly, the leading term at large $n$
is given by (\ref{comm14}), (\ref{comm15}).
Although no explicit factor of $1/n$ is shown in (\ref{comm16}),
we can see that it is down by a power of $n$ because of the factor
$r_{\rm D}^3$ in $dS^{\nu\alpha\beta}$.
One factor $r_{\rm D}$ is used to convert $\del_\nu $ to an angular derivative, while the remaining $r_{\rm D}^2 = M/n$.
We emphasize that despite the simplifications we have carried out, the
result (\ref{comm16}) is still what is derived from the bulk action.
We will show that the quantization of the edge action
as derived in \cite{{KN2},{KN3},{Kar}}, and discussed in Appendix C,
will reproduce 
the result (\ref{comm14}), (\ref{comm15}), so in effect verifying
 the large $n$ limit of (\ref{comm10}). Since the quantization is similar for 
all $\mathbb{CP}^k$, we defer the details of this to the next section.

Finally, turning to the fractional Hall effect of $\nu = 1/m$, by
comparison of the actions (\ref{dolb14b}) and (\ref{dolb16}), we can write
down the anomalous commutation rules (for the Abelian gauge transformations) as
\beqar
[G(\theta^0), G(\theta^{0\prime})] &\equiv& [ J_0 (\theta^0), J_0 (\theta^{0\prime})]\nonumber\\
 &=&
- i { M \over 2 \pi^2 m^2} \oint dS_3\, (\theta^0 \L \theta^{0\prime})\nonumber\\
&&
+ {1\over 8 \pi^2} \oint dS^{\nu\alpha\beta}
(\theta^0 \del_\nu \theta^{0\prime} )\,\left( { {f^0_{\alpha\beta}}\over 3 m^2}
 +{ {R^0_{\alpha\beta}}
\over m}
\right)
\label{comm16b}
\eeqar
The leading term again shows that the action for the edge modes should be
\beq
S = -{M\over 4\pi^2 m^2} \int dt dS_3\,\left[ {\dot \Phi} {\L \Phi} + v_{\rm F} \left({\L \Phi}
\right)^2 \right]
\label{comm9b}
\eeq

\subsection{6+1 and higher dimensions}
Turning to 6+1 dimensions, we see from (\ref{dolb14c})
that one can get contributions from $(CS)_7$, $(CS)_5$ and
$(CS)_3$. While we can simplify the results along lines similar to what we did
above, various expressions become progressively more cumbersome
as we go up in dimensions.
There is however significant simplification in the large $n$ limit.
The dominant term is from $(CS)_3$ with a product of $\Omega$'s, since $\Omega$ is taken to be of order $n$.
From (\ref{dolb14a}), we see that the dominant term in
$2k+1$ dimensions is
\beq
S_{2k +1} =  {1\over (k-1)!} \int  \left( {i \Omega \over 2\pi} \right)^{k-1}
(CS)_3
\label{comm17}
\eeq
Using (\ref{coc11}) and (\ref{comm2b}) we find that the algebra of the
generators of the gauge transformations
works out to be
\beq
[G(\theta), G(\theta')] - i G( \theta\times \theta') =
- {i \over 2 (2\pi)^k} \oint \sqrt{\det i\Omega}~ {\hat x}_\mu (i\Omega^{-1})^{\mu\nu}\,
\Tr (\theta \del_\nu \theta' - \theta'\del_\nu \theta)
\label{comm18}
\eeq
where we have used the relation (\ref{B1}) to carry out the simplification.

Specifically for the case of $\mathbb{CP}^k$, $\Omega = - i n \Omega_{\rm K}$, with
$\sqrt{\det \Omega_{\rm K}} = 2^k /(1+ r^2)^{k+1}$ and this expression can be simplified as
\beqar
[G(\theta), G(\theta')] - i G( \theta\times \theta') &=&
i \,{M^{k-1} \over 2 \pi^k} \oint \Tr (\theta \L \theta') \nonumber\\
&=&- i \,{M^{k-1} \over 2 \pi^k} \oint  \theta^0 ( \L \theta^{0\prime})
\label{comm19}
\eeqar
The second line applies to the Abelian case.
Since $\mathbb{CP}^k = SU(k+1)/ U(k)$, the states of the lowest Landau level
for a symmetric rank-$n$ representation of $SU(k+1)$ which
consists of a number of $SU(k)$ representations, of ranks from zero to
$n$.
As explained in Appendix C, the
droplet is taken to have all the low rank $SU(k)$ representations, 
from rank $= 0$ to rank $= M$, to be occupied \cite{{KN2},{KN3}}.

The effective action for the edge modes 
for an Abelian quantum Hall droplet on $\mathbb{CP}^k$ was derived
in \cite{KN2} as
\beq
S = - {1\over 4 \pi^k} M^{k-1} \int \left[ {\dot \Phi} \L\Phi + v_{\rm F} (\L \Phi )^2
\right] 
\label{comm20}
\eeq
The canonical quantization of this theory is worked out in
Appendix D and gives the commutation rule
\beq
(M^{k-1}/ 2 \pi^k) \oint_y [(\L \theta^0) \Phi(y) , \Phi (x) ] = -i \theta^0(x)
\label{comm21}
\eeq
This shows that we can identify the operator
\beq
J_0 (\theta^0) = -(M^{k-1}/ 2 \pi^k) \oint  \theta^0 \L \Phi
\label{comm22}
\eeq
 as the generator of the Abelian gauge transformations 
$\Phi \rightarrow \Phi + \theta^0$. From the canonical
commutation rule (\ref{comm21}), we also see that
\beq
[ J_0 (\theta^0), J_0(\theta^{0\prime}) ] =  - i\, {M^{k-1} \over 2 \pi^k} \oint \theta^0 \, \L \theta^{0\prime}
\label{comm23}
\eeq
This result is in complete agreement with the commutator anomaly obtained
from topological arguments
in (\ref{comm19}).

As mentioned before, it is possible to have non-Abelian background gauge fields in spatial dimensions $> 2$. There is a corresponding commutator anomaly for the non-Abelian gauge transformations as indicated in
(\ref{comm15}) for $\mathbb{CP}^2$ and in (\ref{comm19})
for the general $\mathbb{CP}^k$. It is possible to compare these results also with the direct quantization of the edge modes.
The action for the edge modes for ${\mathbb{CP}}^k$ 
was obtained by direct calculation
using coadjoint orbits in \cite{KN3}. We found that the edge action is a higher dimensional chiral Wess-Zumino-Witten (WZW) action gauged in a left-right symmetric way with respect to the background non-Abelian gauge field, which is of order $n^0$ and proportional to the $SU(k)$ spin connection of the manifold. The $U(1)$ background field is, as before,
${\bar F} = \Omega = -i n \Omega_{\rm K}$.\footnote{Since we use an antihermitian basis for gauge fields, $A$ in this paper is $i A$ of \cite{Kar}.
Also, to avoid confusion with the generator of gauge transformations,
 we use $V$ in place of $G$ used for the field variable
in \cite{{KN3},{Kar}}.} This construction was further generalized in  \cite{Kar} to include fluctuating gauge fields. This produced a higher dimensional chiral WZW action gauged in a left-right asymmetric way, where $A^{\rm L} = A + \bar{A}$, $A^{\rm R} = \bar{A}$, with $A$, $\bar{A}$ being the fluctuating and fixed background gauge fields, respectively.
The salient features of this effective action are recalled in Appendix C.
Here we quote that it is of the form
\beqar
S &=& {N \over 2 n N'} \biggl[- \int dt d\mu\,
{\del \rho_0 \over \del r^2} \,
\Tr \bigl[ (V^{-1} D_0 V + v_{\rm F} V^{-1} {\hat \L}
V)\, (V^{-1} {\hat \L} V)\bigr] + \Gamma_{\rm WZ} 
\biggr]\nonumber\\
\Gamma_{\rm WZ} &=&
 {k \over 2\pi} \int \rho_0 \, \Tr \biggl[ 
- d \bigl( A^{\rm L} \, d V V^{-1} + A^{\rm R} V^{-1} d V - A^{\rm L} V A^{\rm R}
V^{-1} \bigr)\nonumber\\
&&\hskip .3in  + {1\over 3} ( V^{-1} dV )^3\biggr] \wedge \left( {\Omega_{\rm K}
\over 2\pi}\right)^{k-1}
\label{comm24}\\
{\hat\L} V &=& - 2 r_{\rm D}  {\hat x}_\mu  
(\Omega_{\rm K}^{-1})^{\mu\nu} \left( \del_\nu V + A^{\rm L}_\nu V - 
V A^{\rm R}_\nu \right)\nonumber\\
&=& \L\,V  - 2 r_{\rm D}  {\hat x}_\mu  
(\Omega_{\rm K}^{-1})^{\mu\nu} \left( A^{\rm L}_\nu V - 
V A^{\rm R}_\nu \right)
\label{comm24a}
\eeqar
By virtue of the wedge product with  $(\Omega_{\rm K})^{k-1}$, this is a
K\"ahler-Chern-Simons theory \cite{Nair-Schiff},
a generalization of the WZW model to higher dimensions for a 
K\"ahler manifold. 
Since $\mathbb{CP}^k = SU(k+1)/U(k)$, the
one-particle LLL wave functions $\Psi(u)$ are functions of $u$,
transforming as
an irreducible representation of 
$SU(k+1)$, of dimension $N$ under the left action of
$SU(k+1)$ on $u$, 
and as a representation of dimension $N'$
under the right action of $SU(k) \subset SU(k+1)$ on $u$.
The field $V$ is a unitary $(N' \times N')$-matrix.
We consider the droplet as made of all $SU(k)$ representations
(contained in the $SU(k+1)$ representation corresponding to
$\Psi(u)$) to be filled up to rank $M$.
The density of the droplet is $\rho_0$, which, at large
$n$, is a spherically symmetric step-function,
\beq
\rho_0 = \Theta \bigl( 1- (n r^2 / M) \bigr), \hskip .3in r^2 = \bz\cdot z
\label{comm26}
\eeq

Since the action is of the first order in time-derivatives, and because of
the nature of the WZ term, there are some subtleties in the
canonical quantization. We will give some of the technical details
in Appendix D. The result is that we can define a current
\beq
J(\theta) = {M^{k-1} \over 2\pi^k} \oint dS_{2k-1} \, \Tr [ \theta ({\cal L} V
- V A_\L^{\rm R} ) V^{-1} ]
\label{comm27}
\eeq
where $A_\L^{\rm R} = - 2 r_{\rm D} {\hat x}_{\mu} (\Omega_{\rm K}^{-1})^{\mu\nu} \, A_{\nu}^{\rm R}$ is the angular component of $A^{\rm R}$
corresponding to the direction $\L$.
This current generates left gauge transformations on $V$,
\beq
[ J(\theta), V ] =  -i \,\theta \, V
\label{comm28}
\eeq
The algebra of this current is then 
\beq
[J(\theta), J(\theta') ] = i \, J(\theta \times \theta') + i {M^{k-1} \over 2 \pi^k}
\oint dS_{2k-1} \Tr ( \theta {\cal L } \theta' )
\label{comm29}
\eeq
This is a generalized Kac-Moody algebra.
The WZW action in (\ref{comm24}) is the usual one promoted to
higher dimensions by use of the factor $(\Omega_{\rm K}/2\pi)^{k-1}$.
Effectively we have a set of (1+1)-dimensional chiral theories
on the edge, the spatial direction being the angle Poisson-conjugate to
the radius of the droplet. Therefore the appearance of the Kac-Moody algebra
as in (\ref{comm29}) is basically reproducing, apart from the level number, (\ref{anoalg1}) in (2+1)d and is entirely as expected. 
It is also easy to understand the level number in (\ref{comm29}). 
As shown in Appendix C, it 
is, again as expected, given by the number of chiral modes on the edge of the droplet.

The result (\ref{comm29}) is in complete agreement with the first line of
(\ref{comm19}).
Thus we do have a derivation of the anomalous algebra
of the non-Abelian gauge transformations for the edge modes
from topological arguments as in (\ref{comm19}) and a direct
verification from the edge action in (\ref{comm29}).
We also point out that  actions similar to 
(\ref{comm24}) were derived in \cite{alexios} for fermion droplets in
a general phase space. The results are in agreement for the commutator
anomalies given in (\ref{comm23}) and (\ref{comm29}).

In arriving at the result (\ref{comm29}), we have used the action
(\ref{comm24}), including the gauge fields $A^{\rm L}$, $A^{\rm R}$.
The extension to the algebra turns out to be independent of the gauge fields. This has to do with the fact that in deriving the edge actions we neglected terms of lower order in $1/n$. The inclusion of such terms will lead to 
gauge-field dependent extensions to the algebra similar to the terms
in (\ref{comm16}).

\section{$W_\infty$ and other symmetries}
We now turn to the question of how our results may be viewed as a generalization of the familiar $W_\infty$ symmetry.
Recall that a high value for the background magnetic field
inhibits inter-Landau level transitions, and therefore in quantum Hall systems one can consider dynamics restricted to a specific Landau level,
most often the lowest one. The states of the single particle
Hilbert space are given by the eigenstates of 
a suitable gauged Laplace operator, for which
the background magnetic field
will have 
a component which is Abelian and proportional to 
the $U(1)$ component of the curvature
for a general complex manifold. 
This is the term that can be scaled with $n$ to define a suitable large 
$n$ limit. 
If the manifold is K\"ahler, the Abelian magnetic field
will be proportional
to the K\"ahler form $\Omega_{\rm K}$.
Generally, there can be additional non-Abelian fields as well.

While the Hilbert space is defined by the eigenstates of the Laplacian,
the states of the LLL
also correspond to the states obtained
by quantization of a suitable symplectic two-form.
This is because the projection to the LLL identifies the spatial manifold as the phase space for the dynamics.
For a K\"ahler manifold, the Abelian magnetic field, i.e., 
symplectic form is given by
\beq
\Omega_{\rm symp} = n \, \Omega_{\rm K}
\label{winf1}
\eeq
where $\Omega_{\rm K}$ is the K\"ahler two-form. 
This result applies in all dimensions, the case of the plane is obtained as
the large radius limit of the two-sphere, with $n = 2 B r^2$, for magnetic field $B$.
For a droplet of fermions, the low energy excitations obviously correspond to 
unitary transformations of the single particle Hilbert space
(of the given Landau level). 
Thus, fundamentally, these
are the quantum transformations of interest.
The classical analog of these are canonical transformations preserving the symplectic structure.

Notice that $\Omega_{\rm symp}$ is also the magnetic field, 
so the symplectic potential $\A$ (defined by $\Omega_{\rm symp} = d \A$)
is also the gauge potential for the electromagnetic field.
Thus canonical transformations which preserve $\Omega_{\rm symp}$ can also be viewed as $U(1)$ gauge transformations.

There is one more identification we can obtain in two dimensions.
In this case, $\Omega_{\rm symp}$ is proportional to the area
element, so the canonical transformations can also be viewed as area-preserving diffeomorphisms. These are the $W_\infty$ transformations, whose quantum version is given by the unitary transformations of the single particle Hilbert space. 
Since the droplet of fermions is incompressible due 
to the fermionic nature of the particles, these transformations 
are also naturally
interpreted as incompressible flows of a quantum Hall droplet.

In two dimensions, we can therefore conclude that the algebra of
$W_\infty$ transformations on the edge modes has an extension given by the commutator anomaly for the $U(1)$ gauge transformations.

This is not exactly the state of affairs obtained in higher dimensions.
If the background field is strictly Abelian, then we can use the
symplectic form in (\ref{winf1}). Canonical transformations are then
realized as gauge transformations of the gauge potential. 
The commutator anomaly is still obtained for these transformations acting
on the edge states; however the connection to diffeomorphisms is more subtle.
The natural analog of area-preserving diffeomorphisms in higher dimensions would be volume-preserving diffeomorphisms.
However, there are volume-preserving transformations which do not
preserve $\Omega_{\rm symp}$. In fact any diffeomorphism with
unit Jacobian will preserve the volume.
But these are not acceptable for the dynamics
of the droplet, the acceptable ones must preserve the symplectic two-form
$\Omega_{\rm symp}$. Of course, the acceptable transformations
which preserve $\Omega_{\rm symp}$ will also 
preserve the phase volume $(\Omega_{\rm symp})^k$, which is the geometrically defined volume for a K\"ahler manifold, up to a multiplicative constant. But these are not the most general volume preserving diffeomorphisms. 
Thus, while we do have an anomalous algebra of
transformations on the edge modes, similar to the 2d case, with the anomaly as derived in the previous sections, the transformations are more stringent than the simple analog of area-preserving diffeomorphisms.

There are also additional features in higher dimensions. One can have non-Abelian background fields. In these cases, the single particle Hilbert space
 has a basis of wave functions with an additional index for the non-Abelian symmetry. (In other words, they are sections of a suitable vector bundle, not a line bundle.)
For a proper single particle phase space description, $\Omega_{\rm symp}$
should have additional terms (and additional degrees of freedom)
compared to what is given in (\ref{winf1}).
(This could arise from the action for particles carrying non-Abelian
charge, as in \cite{wong}.)
Nevertheless, we do have the non-Abelian gauge transformations as a symmetry of the full theory, with an algebra for the generators
on edge modes
carrying a commutator anomaly as derived earlier.

We conclude that the Abelian and non-Abelian
gauge transformations restricted to a particular Landau level
are the suitable higher dimensional
generalizations of the $W_\infty$ symmetries in two dimensions.
But these transformations do not have a simple geometric interpretation of
preserving area or volume, or leading to incompressible flows.
They are actually more stringent symmetries.
The anomalous algebra of such transformations on the edge modes is the
natural generalization based on the descent equations of anomalies.

Although we have focused on commutator anomalies for gauge transformations, we note that the bulk actions also have purely gravitational terms
as indicated in (\ref{dolb11}) or like the $\omega d \omega$-term in
(\ref{dolb12}).
They can have associated two-cocycles
which can be viewed as the commutator anomaly for
the generators of the local Lorentz transformations.
For example, in the two-dimensional case, one can have an anomaly
for spatial rotations. 
As is well known, one can also trade the anomaly of Lorentz transformations for a diffeomorphism anomaly \cite{AGG}. In particular, by adding an appropriate local counterterm, the gravitational anomaly may be represented as a local Lorentz anomaly or as a diffeomorphism anomaly. So it may even be appropriate to consider combinations of
these transformations with the gauge transformations, since the latter are related to area-preserving diffeomorphisms.
Clearly there is more to be explored here, we leave this for future work.

\section{Summary}

We have identified the Abelian and non-Abelian gauge transformations restricted to a Landau level as the appropriate higher-dimensional analogs of the $W_{\infty}$ transformations familiar from the two-dimensional quantum Hall effect. In section 2, we gave general arguments for
relating the two-cocycle which gives the extension of the algebra for the generators of gauge transformations to the well-known anomaly descent equations. In effect, this utilizes the
bulk-boundary anomaly inflow. Specifically, we obtained the extended algebra of these transformations, i.e., the higher dimensional
version of the $W_\infty$-algebra, on the edge of a quantum Hall droplet. Explicit formulae for the extended algebras are given  
in section 4 for 2+1, 4+1 and 6+1 dimensions. We emphasize that the key point here is that the construction requires only the bulk topological action and therefore applies even when a complete microscopic edge action is not available. For quantum Hall droplets on $\mathbb{CP}^{k}$, where explicit edge actions are known, we also showed that the resulting extensions agree with direct canonical quantization of the edge modes. This also provides, for cases where it is possible, an independent verification of the general arguments deriving the extended algebra from the bulk action.

A key conclusion of our analysis is that the higher dimensional analogs
of $W_{\infty}$ transformations correspond to transformations which preserve the symplectic two-form. In two dimensions, for dimensional reasons, we could consider them as related to area-preserving diffeomorphisms.
But, in general, there is an important 
distinction compared to the two-dimensional case since the
invariance of the symplectic two-form is a stronger requirement than the preservation of the volume form. Consequently, the higher-dimensional analog of $W_{\infty}$ is not the full algebra of volume-preserving diffeomorphisms,
although the transformations do preserve the volume. Further, higher-dimensional systems admit non-Abelian background fields and corresponding anomalous non-Abelian edge current algebras. They have also been 
derived in section 4.

In the complex geometries studied here, the bulk action is obtained from the Dolbeault index density. However, the anomaly-inflow and descent argument itself is more general. For each context,  it is really a question of what the appropriate index density is. For example, for quantum Hall systems on manifolds without a complex structure, we should expect the relevant index density to be the one for the Dirac complex; its associated Chern-Simons action should
thus play the role of the bulk action leading to the corresponding two-cocycle.
While not discussed in the present paper,
extending the explicit constructions to such manifolds, and incorporating the gauge and gravitational transformations into a unified description, are natural directions for future work.

\bigskip

This work was supported in part by the U.S. National Science Foundation Grants No. PHY-2412479 and PHY-2412480.
\section*{Appendix A: Bulk action from the Dolbeault index theorem}
\def\theequation{A\arabic{equation}}
\setcounter{equation}{0}
We start with a brief review of construction of the bulk effective action 
for $\nu = 1$ for all
dimensions using an appropriate index theorem \cite{KN-dolb}.
The spatial manifold of interest, i.e., $V_{2k}$ discussed
in the last section, will be a complex  manifold $K$ of complex dimension $k$, which includes, as a special case, the complex projective spaces $\mathbb{CP}^k$.

For a $2k$-dimensional real manifold, the spin
connections and curvatures take values in the Lie algebra of the holonomy group $SO(2k)$, corresponding to the group of local frame rotations at a point.
For a complex manifold, coordinate transformations which preserve the complex structure are holomorphic transformations. This restricts the
holonomy group to $U(k) \subset SO(2k)$.
Correspondingly, the frame fields can be taken to be holomorphic and antiholomorphic one-forms (i.e., of the (1,0) and (0,1) type) which are combinations of the real ones
given by the complex structure. The tangent space
also has similar combinations which lead to a complex tangent
space $T_cK$.
The curvature of the manifold takes values in the Lie algebra of $U(k)$.
For defining the Landau levels
on $V_{2k}$, we take the
magnetic fields to be
proportional to the curvatures. It can be abelian, corresponding to the
$U(1)$ generator of $U(k)$, or non-Abelian, taking values in
the Lie algebra of $SU(k) \subset U(k)$.
The single-particle wave functions of the
LLL obey the holomorphicity condition
\begin{equation}
{\bar D}_{\bar i} \Psi = 0
\label{dolb1}
\end{equation}
Thus the wave functions $\Psi$ are part of the Dolbeault complex,
in fact, the twisted Dolbeault complex since we have background gauge fields as well.
The number of normalizable solutions to (\ref{dolb1}) is therefore
given by
the index theorem for the twisted Dolbeault complex as
\begin{equation}
{\rm Index}( \bD ) = \int_K {\rm td}(T_cK) \wedge {\rm ch} (V)
\label{dolb2}
\end{equation}
where ${\rm td}(T_cK)$ is the Todd class on the complex tangent space of $K$
and ${\rm ch}(V)$ is the Chern character of the relevant
vector bundle \cite{eguchi}. 
For any vector bundle with curvature ${\cal F}$, the Chern classes $c_i$
are defined by
\begin{equation}
\det \left( 1 + {i \, {\cal F} \over 2 \pi} \,t\right) = \sum_i c_i \, t^i
\label{dolb3}
\end{equation}
The Todd class may also be represented, via the splitting principle,
in terms of a generating function as
\begin{equation}
{\rm td} = \prod_i {x_i \over 1- e^{-x_i}}
\label{dolb4}
\end{equation}
where $x_i$ represent the ``eigenvalues" of the curvature in a suitable canonical form \cite{eguchi}. This will be the diagonal form for a complex basis
(as in $T_c K$) or 
the canonical antisymmetric form for real antisymmetric $i {\cal F} $. 
Rewriting (\ref{dolb4}) using traces of powers of curvatures, the Todd class can be expanded out as
\begin{equation}
{\rm td} = 1 + {1\over 2} \, c_1 +{1\over 12} ( c_1^2 + c_2) + {1\over 24} c_1\, c_2
+ {1\over 720} ( - c_4 + c_1\,c_3 + 3 \, c_2^2 + 4\, c_1^2 \,c_2 - c_1^4) + \cdots
\label{dolb5}
\end{equation}
The Chern classes involved in this expansion, for low dimensions, 
can be easily worked out from (\ref{dolb3}) and are
given as follows.
\begin{eqnarray}
c_1 (T_c K)& = & \tr ~{iR \over 2\pi} \nonumber \\
c_2 (T_c K)& = & {1 \over 2} \Biggl[ \Bigl(\tr {iR \over 2\pi}\Bigr)^2 - \tr \Bigl({iR \over 2\pi}\Bigr)^2 \Biggr] \nonumber \\
c_3 (T_c K) &=& { 1 \over 3!} \Biggl[ \Bigl(\tr {iR \over 2\pi}\Bigr)^3 - 3\, \tr {iR \over 2\pi} \,\tr \Bigl({iR \over 2\pi}\Bigr)^2 + 2 \,\tr \Bigl({iR \over 2\pi}\Bigr)^3 \Biggr] \\
c_4 (T_cK) &=& { 1 \over 4!} \Biggl[ \Bigl(\tr {iR \over 2\pi}\Bigr)^4 - 6 \Bigl(\tr {iR \over 2\pi}\Bigr)^2 \,\tr \Bigl({iR \over 2\pi}\Bigr)^2 + 8\, \tr {iR \over 2\pi}\, \,\tr \Bigl({iR \over 2\pi}\Bigr)^3\nonumber\\
&&\hskip .4in  +3\, \tr \Bigl({iR \over 2\pi}\Bigr)^2 
\,\tr \Bigl({iR \over 2\pi}\Bigr)^2 -6\,\tr \Bigl({iR \over 2\pi}\Bigr)^4 \Biggr] \nonumber
\label{dolb6}
\end{eqnarray}

The vector bundle $V$ relevant to the Chern character in (\ref{dolb2}),  is defined by the internal gauge symmetry structure of the fermions fields. The fermion
wave functions are sections of this bundle.
The Chern character can be written out as
\begin{equation}
{\rm ch}(V) = \tr \left( e^{i {\cal F} /2 \pi} \right) = {\rm dim}\,V + \Tr ~{{i {\cal F}} \over {2\pi}} + { 1 \over 2!} \Tr~ {{i{\cal F} \wedge i {\cal F}} \over {(2\pi)^2}} + \cdots
\label{dolb7}
\end{equation}
where ${\rm dim} V$ is the dimension of the bundle $V$.
Here ${\cal F}$ is the gauge field strength $F$.
(If spin is included,  ${\cal F}$ will also
include the curvature of the spin bundle.)

Finally, we note that the integral in the index theorem (\ref{dolb2}) is to be interpreted in the standard way,
namely, in the product of the Todd class and the Chern character, we pick the 
set of terms which give a $2k$-form and integrate it over the manifold.

The argument for the construction of the bulk action in \cite{KN-dolb} is 
the following.
The index gives the degeneracy of the LLL. Therefore the charge density
$J_0$, for fermions of unit charge, can be identified with the index density
for the $\nu =1$ state. In terms of the effective action, we can thus write
\begin{equation}
{\delta S^{\rm bulk} \over \delta A_0} = J_0 =  {\rm Index ~density}
\label{dolb8}
\end{equation}
This shows that the leading term of the effective action may be taken as
a Chern-Simons term $CS(A)$ whose variational derivative with respect to
$A_0$ will give the index density. In other words, we can
``integrate up" the relation (\ref{dolb8}), and appropriately covariantize to obtain the topological part of the bulk effective action
$S$.
(In principle, there can be terms arising from
dipole and higher multipole terms in $J_0$. These 
do not affect the degeneracy, but can contribute to the action.
However, because they involve multipole terms in $J_0$, they will lead to
higher derivatives of the fields and will be metric dependent as well.
The action given by the index density
will be the topological term and is also
the leading term in the sense of a derivative expansion.)

This argument
does not determine the purely gravitational terms in
the effective action.
If we have a droplet with edge modes, they can generate
a gravitational anomaly as well. 
The purely gravitational terms in $S$ should be determined 
by the gravitational anomaly via the descent method used for anomalies. 
For more details on this, see\cite{KN-dolb}.

The conclusion is that the bulk effective action for $\nu =1$ higher dimensional
QHE is of the form
\begin{equation}
S^{\rm bulk} =
 \int \Bigl[ {\rm td}(T_c K) \wedge \sum_p  (CS)_{2 p+1} ( A)\Bigr]_{2 k+1}
+ 2 \pi\int \Omega^{\rm grav}_{2k+1}
\label{dolb9}
\end{equation}
Here $(CS)_{2 p +1}(A)$ is the Chern-Simons term associated with
just the gauge part and is defined by
\begin{equation}
{1\over 2 \pi} d (CS)_{2 p +1} = {1 \over (p+1)!} \Tr \left(
{ i F \over 2 \pi}\right)^{p +1}
\label{dolb10}
\end{equation}
In terms of $\omega_{2p+1}$, we have $(CS)_{2 p +1}  =  \omega_{2p+1}$, with $d\omega_{2p+1}$ as the purely gauge field dependent part of the index density. 
Again, as in the case of the index theorem, one should 
expand the terms in the square brackets in
(\ref{dolb9}) in powers of
curvatures and $F$ and pick out the terms corresponding to
the $(2k+1)$-form. This is indicated by the subscript $2k+1$ for
the square brackets.

The purely gravitational contribution is the last term in
(\ref{dolb9}). It is the analogue of the Chern-Simons term
for the Todd class defined by
\beq
d \Omega^{\rm grav}_{2k+1} = [{\rm td}(T_c K)]_{2k+2}
\label{dolb10a}
\eeq
The argument for the term $S^{\rm grav}$ is that
the gravitational anomaly for the $2k$-dimensional 
boundary field theory is obtained from the $(2k+2)$-dimensional
index density by the standard descent procedure.
\section*{Appendix B: Some useful formulae}
\def\theequation{B\arabic{equation}}
\setcounter{equation}{0}
In this Appendix, we collect some formulae useful for simplifications carried out in text.
In general, for an antisymmetric matrix $M_{\mu\nu}$, we have
\beqar
\epsilon^{\alpha_1\alpha_2\alpha_3\alpha_4\cdots \alpha_{2k-1}\alpha_{2k} }M_{\alpha_1\alpha_2} M_{\alpha_3\alpha_4} \cdots M_{\alpha_{2k-1}\alpha_{2k}}
&=&2^k k! \sqrt{\det M}\nonumber\\
\epsilon^{\mu\nu\alpha_3\alpha_4\cdots \alpha_{2k-1}\alpha_{2k} }M_{\alpha_3\alpha_4} \cdots M_{\alpha_{2k-1}\alpha_{2k}} &=& 
2^k k! \left[ - {1\over 2 k}\right] (M^{-1})^{\mu\nu} \sqrt{\det M}
\label{B1}
\eeqar
We have used these relations in sections 4.2, 4.3 in text, setting
$M_{\mu\nu} = i \Omega_{\mu\nu}$.
Specifically for the manifold $\mathbb{CP}^k$, the K\"ahler form is the
Fubini-Study form given in (\ref{comm11}).
This obeys the integral formula
\beq
\int_{\mathbb{CP}^k} \left( {\Omega_{\rm K} \over 2\pi}\right)^k 
= 1
\label{B2}
\eeq
The metric on $\mathbb{CP}^k$ is the Fubini-Study metric given by
\beq
ds^2 =  \left[ {dz^i \, d\bz^i \over (1+ \bz \cdot z)}
- { \bz\cdot dz\, z\cdot d\bz \over (1+ \bz \cdot z)^2}
\right] 
\label{B3}
\eeq
It is the symmetric product of the differentials in (\ref{B3}), there is
no wedge product. In terms of real coordinates $x^{\mu}, \mu= 1,\cdots , 2k$, defined by
$z^i = x^{2i -1} + i x^{2i}$, $\bz^i = x^{2i -1} - i x^{2i}$,
$i= 1, 2,\cdots, k$, the
volume element corresponding to this metric is
\beq
d \mu = {k! \over \pi^k} {d^{2k} x \over (1+ r^2)^{k+1}},
\hskip .3in
\int_{\mathbb{CP}^k} d\mu = \int_{\mathbb{CP}^k} \left( {\Omega_{\rm K} \over 2\pi}\right)^k = 1
\label{B4}
\eeq
where $r^2 = x^\mu x_\mu$.
This is the volume element which appears in the first term of
(\ref{comm24}).
For the case of $\mathbb{CP}^2$, these relations give
\beqar
\epsilon^{\mu\nu\alpha\beta} (\Omega_{\rm K})_{\mu\nu} (\Omega_{\rm K})_{\alpha\beta}
&=& 8 \sqrt{\det \Omega_{\rm K}} = {32 \over (1+ r^2)^3}, \hskip .3in r^2 = x^\mu x_\mu\nonumber\\
\epsilon^{\mu\nu\alpha\beta} (\Omega_{\rm K})_{\mu\nu}&=&
- 2 \sqrt{\det \Omega_{\rm K}} \, (\Omega_{\rm K}^{-1})^{\alpha\beta}
\label{B5}
\eeqar

\section*{Appendix C: Effective adge actions for QHE on $\mathbb{CP}^k$}
\def\theequation{C\arabic{equation}}
\setcounter{equation}{0}
The effective edge actions for $\nu=1$ QHE on $\mathbb{CP}^k$, both with constant Abelian and non-Abelian gauge fields were derived in \cite{{KN2},{KN3}} using a coadjoint orbit construction. Here we summarize the results. 

\noindent{\bf 1. Abelian case}

The states of the lowest Landau level for $\mathbb{CP}^k$ form a symmetric representation of $SU(k+1)$ of rank $n$, which may be decomposed into a series of $SU(k)$ representations. For simplicity we have considered a spherical droplet by having the first $M$ $SU(k)$ representations occupied. 
To obtain such a droplet as the multifermion ground state 
we used an appropriate spherically symmetric confining potential.
The density of the droplet, in the large $n$ limit, then has the form
\beq
\rho_0 = \Theta \bigl( 1- (n r^2 / M) \bigr), \hskip .3in r^2 = \bz\cdot z
\label{D5}
\eeq
This defines a spherical droplet of radius $r_{\rm D}$, where $r_{\rm D}^2 = M/n$ for large $n$ and $M$, with $n \gg M$.

For an Abelian constant magnetic field of the form $\bar{F}= -in \Omega_{\rm K}$, where $\Omega_{\rm K}$ is the Fubini-Study K\"ahler form defined in (\ref{comm11}), we found that, in the large $n$-limit, the quantum Hall dynamics is confined to the boundary of the droplet and is described by a chiral action of the form 
\beq
S = - {1\over 4 \pi^k} M^{k-1} \int dt dS^{2k-1} \left[ {\dot \Phi} \L\Phi + v_{\rm F} (\L \Phi )^2
\right] 
\label{D1}
\eeq
where $\Phi$ is a scalar field, $v_{\rm F}$ is the Fermi velocity 
(which is related to the gradient of the confining potential energy
on the edge of the droplet) and $\L$ defines a particular tangential direction along the boundary of the droplet. In particular, 
\beq
\L =- 2\, r_{\rm D}\, {\hat x}_{\mu} (\Omega_{\rm K}^{-1})^{\mu\nu}
\del_{\nu} = 
 i \left( z\cdot {\del \over \del z} - \bz \cdot {\del \over \del \bz}
\right)
\label{D2}
\eeq

In two dimensions, with $\mathbb{CP}^1 \sim S^2$, $\L = {\del \over \del \vf}$ is the derivative tangential to the circular edge boundary and the equation of motion derived from (\ref{D1}) shows the chirality of the edge modes. In higher dimensions, there are several tangential directions along the boundary droplet. However, the theory selects a particular tangential direction $\L$ for the propagation of the edge modes. This is Poisson conjugate to the radial direction $\hat{x}^{\mu}$, as indicated in (\ref{D2}). Effectively, the higher dimensional edge action is a collection of $(1 +1)$-dimensional chiral edge modes. The number of such modes is indicated by the overall level number in (\ref{D1}). In particular, since $S^{2k-1}/S^1 = \mathbb{CP}^{k-1}$,
separating out the $S^1$ direction (corresponding to $\L$), we get a
$\mathbb{CP}^{k-1}$-worth of (1+1)-dimensional theories
 \cite{{KN2},{KN3}}.
The volume of $\mathbb{CP}^{k-1}$ as given by the integral over
$dS_{2k-1}$, after factoring out $S^1$, is $\pi^{k-1} /(k-1)!$.
The edge modes which are the last occupied $SU(k)$ states form a rank
$M$ symmetric representation, which is of dimension 
\beq
{(M+k -1)!\over (M! (k-1)!) } \approx {M^{k-1}\over  (k-1)! }
= {\pi^{k-1} \over (k-1)!} \left[ {M^{k-1} \over  \pi^{k-1}}\right]
\label{D2a}
\eeq
for large $M$.
Since the factor $\pi^{k-1} /(k-1)!$ is already accounted for by
$dS_{2k-1}$ in (\ref{D1}), we see that
the level number is correctly given by the number of chiral modes
of the edge of the droplet.

\noindent{\bf 2. Non-Abelian case}

The formulation of QHE on $\mathbb{CP}^{k}$ allows the possibility of non-Abelian background magnetic fields. In particular, one can choose ``constant" magnetic fields proportional to the $U(k)$ curvature components. This means that, in addition to the background
$U(1)$ magnetic field
which is given by $\Omega$, we choose
the $SU(k)$ gauge potential ${\bar A}$ proportional to the
spin connection of $\mathbb{CP}^k$ \cite{KN3}.
 In this case, the single-particle lowest Landau level wave functions are sections of a $U(k)$ bundle on $\mathbb{CP}^{k}$ carrying appropriate $U(k)$ charges. In particular, since $\mathbb{CP}^k = SU(k+1)/U(k)$ can be parametrized by
a group element $u \in SU(k+1)$, the
one-particle LLL wave functions $\Psi(u)$ are functions of $u$.
They transform as
an irreducible representation of 
$SU(k+1)$, of dimension $N$ under the left action of
$SU(k+1)$ on $u$, 
and as a representation of dimension $N'$
under the right action of $SU(k) \subset SU(k+1)$ on $u$. (In the Abelian case, $N'=1$, and the wavefunctions are singlets under the right action of $SU(k)$.)

The $SU(k+1)$ representation of the LLL states
can be viewed as made of a set of
$SU(k)$ representations. All such $SU(k)$ representations up to rank $M$
are taken to be filled to form the droplet.
The density of the droplet is $\rho_0$, which, at large
$n$, is again a spherically symmetric step-function as in (\ref{D5}).
For the large $n$ simplification of the action, we can again use
$N/N' \rightarrow n^k/k!$.

We found in \cite{KN3} that the effective action at large $n$ is also a chiral boundary action which has the form of a generalized, higher dimensional 
Wess-Zumino-Witten action given by
\beqar
S
&=& -{N \over {2nN'}} \int  dt d\mu ~{{\del \rho_0} \over \del r^2} ~
 \tr \left[ \left( V^{-1} {\dot V} + v_{\rm F} ~V^{-1} {\hat\L} V \right)  
V^{\dagger}{\hat\L} V  \right] \nonumber \\
&&+ {{Nk} \over { 4 \pi nN'}} \int \rho_0 \left[ -d \left( i \bar{A} dV V^{-1} + i \bar{A} V^{-1}dV \right) + {1 \over 3} \left( V^{-1}dV \right)^3 \right] \wedge \left( {\Omega_{\rm K} \over {2 \pi}}\right)^{k-1} 
\label{D3}
\eeqar
where the field $V$ is a unitary $(N'\times N')$-matrix\footnote{ In earlier papers, this field was denoted by G. We have changed the notation here to avoid confusion, as G is used to denote a different quantity in the present paper.} 
and 
\beq
{\hat\L} = - (\Omega_{\rm K}^{-1})^{\mu\nu} 2\, r_{\rm D} \hat{x}_\mu D_\nu
\label{D4}
\eeq
${\hat\L}$ is the component of the covariant derivative $D= \del + [\bar{A}, ~]$ perpendicular to the radial direction, along a special tangential direction on the droplet boundary. This is the gauged version of the operator
$\L$ defined in (\ref{D2}). The density of the droplet is the spherically
symmetric $\rho_0$
given in (\ref{D5}).

The action $S$ in (\ref{D3}) is a higher dimensional generalization of a Wess-Zumino-Witten action, vectorially gauged with respect to the time independent background gauge field $\bar{A}$. The equations of motion further show that the propagation is along a particular direction and the theory is chiral.  
The first term on the right hand side in (\ref{D3}) is evidently 
on the boundary of the droplet since the derivative of $\rho_0$ is a $\delta$-function. The second term is a WZW-type term written as  an integral over a manifold of dimension equal to $(2k+1)$, corresponding to the droplet and time. The usual 3-form in the integrand of the WZW-term, $(V^{-1} d V)^3$, has now been augmented to the appropriate $(2k+1)$-form $(V^{-1} d V)^3 \wedge \Omega_{\rm K}^{k-1}$. Since the WZW-term is the integral of a locally exact form \cite{KN3}, the whole action $S$ is an edge action. 

Further, as in the Abelian case, because the propagation is along a particular tangential direction, namely the Poisson conjugate to the radial direction, the higher dimensional edge action (\ref{D3}) can be considered as a collection of (1+1)-dimensional non-Abelian chiral modes. Since $N/N' \rightarrow n^k/k!$ for large $n$, one can easily show that 
the multiplicity and the level number are the same as in the Abelian case.  (The action (\ref{D3}) coincides with (\ref{D1}) when $N'=1$). 

Once we allow for fluctuating (dynamical) gauge fields,
the chiral action (\ref{D3}) becomes anomalous. It was shown in \cite{Kar} that the effective action in the presence of fluctuating gauge fields is of the form
\beqar
S
&=& -{N \over {2nN'}} \int  dt d\mu ~{{\del \rho_0} \over \del r^2} ~
 \tr \left[ \left( V^{-1} {D_0 V} + v_{\rm F} ~V^{-1} {\hat \L} V \right)  
V^{-1} {\hat \L} V  \right]  \label{D6} \\
&&+ {{Nk} \over { 4 \pi nN'}} \int \rho_0 \left[ -d \left(  {A^{\rm L}} dV V^{\dagger} +  {A^{\rm R}} V^{\dagger}dV - A^{\rm L} V A^{\rm R} V^{-1} \right) + {1 \over 3} \left( V^{-1}dV \right)^3 \right] \wedge \left( {\Omega_{\rm K} \over {2 \pi}}\right)^{k-1} 
\nonumber
\eeqar
where $A^{\rm L} = A + \bar{A},~~A^{\rm R} = \bar{A}$, with $A,~\bar{A}$ being the fluctuating and and fixed background gauge field, respectively. The derivative operator ${\hat \L}$, as before, denotes the component of the generalized covariant derivative $D_\mu V = \del_\mu V + A_\mu^{\rm L} V - V A_\mu^{\rm R}$ along the tangential direction Poisson conjugate to the
radius, i.e.,
\beq
{\hat\L} V = - 2 r_{\rm D}  {\hat x}_\mu  
(\Omega_{\rm K}^{-1})^{\mu\nu} \left( \del_\nu V + A^{\rm L}_\nu V - 
V A^{\rm R}_\nu \right)
\label{D7}
\eeq

The effective edge action (\ref{D6}) is a higher dimensional chiral Wess-Zumino-Witten action gauged in a left-right asymmetric way. 
Gauge transformations correspond to left transformations on $V$,
which would be, for an invariant action, compensated by the transformation of the fluctuating field $A$.
The action (\ref{D6}) is however not gauge-invariant; its
non-invariance under gauge transformations is cancelled by the 
change in the bulk topological action \cite{Kar}.

\section*{Appendix D: Canonical Quantization of the Edge Modes}
\def\theequation{D\arabic{equation}}
\setcounter{equation}{0}
The quantum dynamics of the edge modes 
can be obtained using standard methods of geometric
quantization. We will start with the Abelian case where the kinetic term in the action is of the form
\beq
S = C \int dt \oint\, {\dot \Phi} \, {\cal L} \Phi
\label{C1}
\eeq
where $C$ is a constant.
We consider an interval of time $[t_{\rm i}, t_{\rm f}]$.
The variation of the action takes the form
\beq
\delta S =  -2 C \int dt \oint~ \delta \Phi\, ( {\cal L} {\dot \Phi})
+ C\,\oint\,  {\cal L} \Phi\, \delta \Phi \Big\vert^{t_{\rm f}}_{t_{\rm i}} 
\label{C2}
\eeq
This identifies the canonical (or symplectic) one-form  and two-form for the theory as
\beq
\A = C \oint \,  {\cal L} \Phi\, \delta \Phi, \hskip .3in
\Omega_{\rm symp} =\delta \A = C \oint \, ({\cal L}\delta \Phi) \, \delta \Phi
\label{C3}
\eeq
where $\delta$ denotes the exterior derivative on the space of fields.

Consider the vector field 
$V_\theta = \oint (\theta^0\, {\delta/\delta \Phi})$. The interior contraction of this with $\Omega_{\rm symp}$ is 
\beq
V_\theta \rfloor \Omega_{\rm symp} =
2 C \oint \, ({\cal L} \theta^0) \, \delta \Phi
= - \delta \left[ - 2 C \oint \, ({\cal L} \theta^0 ) \Phi \right]
\label{C4}
\eeq
This identifies $J_0 (\theta^0) = - 2C  \oint \, ({\cal L} \theta^0 ) \Phi $ as the function
on the phase space which, via Poisson brackets, implements
the transformation given by $V_\theta$.
Using $\{ f, h\} = - V_f \rfloor \delta h$, we can work out the Poisson brackets
to get
\beq
\{ - 2C  \oint \, ({\cal L}\theta^0 ) \Phi , \Phi (x)\}
= - V_\theta\rfloor \delta \Phi (x) = - \theta^0 (x)
\label{C5}
\eeq
We can now obtain the quantum commutation rules
\beqar
[J_0 (\theta^0), \Phi (x) ] &=& - i \, \theta^0(x)\nonumber\\
{}[ J_0 (\theta^0), J_0 (\theta'^0) ] &=& i\, \bigl( - V_\theta\rfloor \delta J_0 (\theta'^{0})\bigr)
= i \, 2 C \oint \, \theta^0 ({\cal L} \theta'^0)
\label{C6}
\eeqar
This leads to the results in (\ref{comm8}), (\ref{comm23})
in text, with $C = - M^{k-1} /(4 \pi^k)$.

In the case of the non-Abelian edge modes with the action (\ref{comm24}), we
can follow a similar strategy of varying the action and carrying out an integration-by-parts for the terms with the time-derivative of the field
variations. The boundary term at $t_{\rm f}$ will give the canonical one-form
$\A$. For the action (\ref{comm24}), we find
\beqar
\A &=&\A_1 + \A_2 + \A_3 + \A_4\nonumber\\
\A_1 &=&\lambda \int d \mu\, \del_\mu \rho_0\, (\Omega_{\rm K}^{-1})^{\mu\nu}
\, \Tr \left( V^{-1} \delta V\, V^{-1} D_\nu V\right)\label{C7}\\
\A_2&=& -{\lambda k \over 2\pi}\int \rho_0~ 
d \left( \Tr \left[ \delta V\, V^{-1} A^{\rm L} + A^{\rm R} \,V^{-1} \delta V\right]
\right) \wedge \left( {\Omega_{\rm K} \over 2\pi}\right)^{k-1}
\label{C8}\\
\A_3 &=& {\lambda k \over 2\pi}\int \rho_0 ~ d \, \left(\Tr \left[ 
\delta V \, V^{-1} dV \, V^{-1} \right] \right)\wedge \left( {\Omega_{\rm K} \over 2\pi}\right)^{k-1}\label{C9}\\
\A_4 &=&- {\lambda k \over 2\pi}\int \rho_0 ~
\Tr \left[ d (\delta V \, V^{-1}) \, d V \, V^{-1} \right] \wedge
\left( {\Omega_{\rm K} \over 2\pi}\right)^{k-1}
\label{C10}
\eeqar
where $\lambda = N/(2 n N')$.
A couple of comments are in order. $\A_1$ arises from the kinetic term
in (\ref{comm24}), $\A_2$, $\A_3$ and $\A_4$ are due to
$\Gamma_{\rm WZ}$. The variation of
$\Gamma_{\rm WZ}$ yields a total derivative.
The time-derivative part of it integrates to a boundary term at
$t_{\rm f}$ with no further time-derivatives. 
This is the term $\A_4$ in (\ref{C10}). The appearance of this term is
tied to the fact that we cannot express $\Gamma_{\rm WZ}$ entirely 
in terms of $2k$-dimensional spacetime integral of a local density.
An extra coordinate is needed, which, in the present case is the
radial coordinate of the droplet.
The expression for $\delta \A_4$ will produce a total derivative and
so, while $\A_4$ cannot be written as an expression on the boundary
of the droplet, its contribution to $\Omega_{\rm symp}$ is
defined on the boundary.
There are also terms in $\delta \Gamma_{\rm WZ}$ which are on the boundary of the
droplet, which, upon a further integration over time, gives the additional
terms $\A_2$, $\A_3$.
We will shortly convert these terms to integrals on the boundary
of the droplet, but, for now, we have displayed them as
integrals of total (spatial) derivatives.
Using the expression for $\rho_0$ from (\ref{comm26}), we can 
further simplify all these terms. For $\A_1$ we get
\beqar
\A_1 &=& {\lambda k! \over \pi^k} \int dS_{2k-1} dr {r^{2k-1} \over (1+ r^2)^{k+1}} \left[ - \delta \left( {M \over n}- r^2\right) \right] 2 x_\mu (\Omega_{\rm K}^{-1})^{\mu\nu} \Tr \left( V^{-1} \delta V \, V^{-1} D_\nu V\right)
\nonumber\\
&\approx&  {\lambda k! \over 2 \pi^k}{M^{k-1} \over n^{k-1}}
 \int dS_{2k-1}  \Tr \left( V^{-1} \delta V \, V^{-1} {\hat\L} V \right)
 \label{C11}
 \eeqar
 where
${\hat\L} V$, as given in (\ref{D7}), 
is the gauged version of the derivative $\L$ given in
(\ref{comm13a}). Upon taking a second variation, we find
\beq
\Omega_1 \equiv \delta \A_1 =
 {\lambda k! \over 2 \pi^k}{M^{k-1} \over n^{k-1}}
 \int dS_{2k-1}  \Tr \left( v^2  \, V^{-1} {\hat\L} V
 - v\, ({\hat\L} v) \right)
 \label{C13}
\eeq
where $v = V^{-1} \delta V$ and 
${\hat\L} v = \L v + A^{\rm R} v - v A^{\rm R}$.
Notice that $v = V^{-1} \delta V$ is invariant under group translations
on the left of $V$ and undergoes conjugation for the right translations.
So ${\hat\L} v$ is the correct gauge covariant derivative of $v$.

For the remaining terms (\ref{C8})-(\ref{C10}) it is easier to take the second
variation before simplifying. We find
\beqar
\delta \A_2&=& - {\lambda k \over 2\pi} 
\int \rho_0\, d\left[ \Tr \left( v^2 ( V^{-1} A^{\rm L} V - A^{\rm R} )
\right) \right]\wedge \left( {\Omega_{\rm K} \over 2\pi}\right)^{k-1}
\nonumber\\
\delta \A_3 &=& {\lambda k \over 2\pi} 
\int \rho_0 \, d\left[ \Tr \left( - v d v + v^2\, V^{-1} dV \right) \right]
\wedge \left( {\Omega_{\rm K} \over 2\pi}\right)^{k-1}
\label{C14}\\
\delta\A_4&=&-{\lambda k \over 2\pi} \int \rho_0 \, d\left[\Tr \left( - v dv
+ 2 v^2 \, V^{-1} dV\right) \right] \wedge \left( {\Omega_{\rm K} \over 2\pi}\right)^{k-1}
\nonumber
\eeqar
Combining these terms we get
\beqar
\delta ( \A_2 + \A_3 +\A_4)
&=& -{\lambda k \over 2\pi} \int \rho_0 \,  d \left( \Tr \left[ v^2 \, V^{-1} D V 
\right] \right) \wedge \left( {\Omega_{\rm K} \over 2\pi}\right)^{k-1}
\nonumber\\
&=& - {\lambda k! \over 2 \pi^k} {M^{k-1} \over n^{k-1}}
\int dS_{2k-1} \, \Tr \left( v^2 \, V^{-1} {\hat\L} V \right)
\label{C15}
\eeqar
We now add this to (\ref{C13}) to get
\beqar
\Omega_{\rm symp} &=&- {\lambda k! \over 2 \pi^k} {M^{k-1} \over n^{k-1}}
\int dS_{2k-1} \,\Tr \bigl( v ({\hat\L} v) \bigr)\nonumber\\
&\approx& - {M^{k-1}\over 4 \pi^k} 
\int dS_{2k-1} \, \Tr \bigl( v ({\hat \L} v)\bigr)
\label{C16}
\eeqar

Once again we can consider left translations on $V$, with the vector
field $V_\theta$ such that $V_\theta \rfloor \delta V = \theta\, V$.
The contraction of this with $\Omega_{\rm symp} $ gives
\beq
V_\theta \rfloor \Omega_{\rm symp} 
= - \delta \left[ {M^{k-1} \over 2 \pi^k} \int dS_{2k-1} \Tr \left( \theta ({\L} V
- V A_\L^{\rm R} ) V^{-1} \right)\right]
\label{C17}
\eeq
where $A_\L^{\rm R} = - 2 r_{\rm D} {\hat x}_\mu (\Omega_{\rm K}^{-1})^{\mu\nu} \, A_\nu^{\rm R}$.
We therefore define the generator of the transformation as
\beq
J(\theta) =  {M^{k-1} \over 2 \pi^k} \int dS_{2k-1} \Tr \left( \theta ({\L} V
- V A_\L^{\rm R} ) V^{-1} \right)
\label{C18}
\eeq
Working out the Poisson brackets and upgrading to the commutation rules, we find the algebra
\beqar
[J(\theta), V ]&=& -i\, \theta \, V\nonumber\\
{}[J(\theta), J(\theta') ] &=& i J(\theta \times \theta') +
i {M^{k-1} \over 2 \pi^k} \int dS_{2k-1}\, \Tr \bigl( \theta (\L \theta')\bigr)
\label{C19}
\eeqar
These are the results (\ref{comm28}) and (\ref{comm29}) quoted in
text.


\end{document}